%% file: mainfile.tex
\documentclass[vecphys]{svmult}
\usepackage{makeidx}         		% allows index generation
\usepackage{graphicx}        		% standard LaTeX graphics tool
\usepackage{multicol}        		% used for the two-column index
\usepackage[bottom]{footmisc}		% places footnotes at page bottom
\usepackage{graphicx,epic,eepic}
\usepackage[dvips]{color}
\usepackage{amsmath,amsfonts,amssymb}
\makeindex  % used for the subject index
\newcommand{\betrag}[1]{\left\vert#1\right\vert}

\newcommand{\MatrixB}[4]{\left[ \begin{array}{cc} #1 & #2 \\ #3 & #4 \end{array}\right]}
\def\diag{{\mathrm{diag}}}			% Diagonal matrix
\def\det{{\mathrm{det}}}			% Determinant
\def\tr{{\mathrm{tr}}} 				% Trace
\def\T{{\mathrm{T}}} 				% Transpose

\def\ipd{{\mathrm{\Theta}}}        % short for initial phase difference
\def\K{{\mathrm{K}}}
\def\G{{\mathrm{\Gamma}}}

\def\Pl{{\mathrm{Planck}}}           	% Planck
\def\A{{\mathrm A}} 				% Condensate A
\def\B{{\mathrm B}} 				% Condensate B

\def\Rb{{\mathrm{Rb}}} 			% Rubidium Atoms
\def\EFT{EFT}					% effective field theory
\def\EFTs{EFTs}				% effective field theoies

\def\d{{\mathrm{d}}} 				% d in dx for example
\def\r{{\mathbf{r}}}

\def\ie{{\emph{i.e.}}}				% i.e. for the text
\begin{document}
%%%%%%%%%%%%%%%%%%%%%%%%%%%%%%%%%%%%%%%%%%%%%%%%%%%
\title*{Analogue spacetime based on 2-component Bose--Einstein condensates}
\author{Silke Weinfurtner\inst{1}\and
Stefano Liberati\inst{2}\and
Matt Visser\inst{3}}
\institute{Victoria University\\ 
School of Mathematics, Statistics, and Computer Sciences\\ 
PO Box 600, Wellington, New Zealand\\ 
E-mail: \texttt{silke.weinfurtner@mcs.vuw.ac.nz}
\and International School for Advanced Studies\\ 
Via Beirut 2-4, 34014 Trieste, Italy and INFN, Trieste\\ 
E-mail: \texttt{liberati@sissa.it}
\and Victoria University\\ 
School of Mathematics, Statistics, and Computer Sciences\\ 
PO Box 600, Wellington, New Zealand\\ 
E-mail: \texttt{matt.visser@mcs.vuw.ac.nz}}
%%%%%%%%%%%%%%%%%%%%%%%%%%%%%%%%%%%%%%%%%%%%%%%%%%%
\maketitle
%%%%%%%%%%%%%%%%%%%%%%%%%%%%%%%%%%%%%%%%%%%%%%%%%%%
\begin{abstract}
  Analogue spacetimes are powerful models for probing the fundamental
  physical aspects of geometry --- while one is most typically
  interested in ultimately reproducing the pseudo--Riemannian
  geometries of interest in general relativity and cosmology, analogue
  models can also provide useful physical probes of more general
  geometries such as pseudo--Finsler spacetimes. In this chapter we
  shall see how a 2-component Bose--Einstein condensate can be used to
  model a specific class of pseudo--Finsler geometries, and after
  suitable tuning of parameters, both bi-metric pseudo--Riemannian
  geometries and standard single metric pseudo--Riemannian geometries,
  while independently allowing the quasi-particle excitations to
  exhibit a ``mass''.  Furthermore, when extrapolated to extremely
  high energy the quasi-particles eventually leave the phononic regime
  and begin to act like free bosons. Thus this analogue spacetime
  exhibits an analogue of the ``Lorentz violation'' that is now
  commonly believed to occur at or near the Planck scale defined by
  the interplay between quantum physics and gravitational physics. In
  the 2-component Bose--Einstein analogue spacetime we will show that
  the mass generating mechanism for the quasi-particles is related to
  the size of the Lorentz violations. This relates the ``mass
  hierarchy'' to the so-called ``naturalness problem''.  In short the
  analogue spacetime based on 2-component Bose--Einstein condensates
  exhibits a very rich mathematical and physical structure that can be
  used to investigate many issues of interest to the high-energy
  physics, cosmology, and general relativity communities.
\end{abstract}

%%%%%%%%%%%%%%%%%%%%%%%%%%%%%%%%%%%%%%%%%%%%%%%%%%	                                    

\section{Introduction and motivation\label{Intro}}						

%%%%%%%%%%%%%%%%%%%%%%%%%%%%%%%%%%%%%%%%%%%%%%%%%%%

Analogue \index{analogue~spacetime} models of curved spacetime are
interesting for a number of reasons~\cite{LRR}: Sometimes the analogue
spacetime helps us understand an aspect of general
relativity,\index{general~relativity} sometimes general relativity
helps us understand the physics of the analogue spacetime, and
sometimes we encounter somewhat unusual mathematical structures not
normally part of the physics mainstream, with the payoff that one
might now develop new opportunities for exploiting the traditional
cross-fertilization between theoretical physics and
mathematics~\cite{normal,normal2,birefringence}.

Specifically, in this chapter we will discuss an analogue
spacetime\index{analogue~spacetime} based on the propagation of
excitations in a 2-component
Bose--Einstein\index{Bose--Einstein~condensate}
condensate (BEC)~\cite{mass1,mass2,qgp1,qgp2,qgp3,qgp4}. This analogue
spacetime has a very rich and complex structure. In certain portions
of parameter\index{parameter~space} space the most natural
interpretation of the geometry is in terms of a specific class of
pseudo--Finsler \index{pseudo--Finsler~spacetime} spacetimes, and
indeed we will see how more generally it is possible to associate a
pseudo--Finsler \index{pseudo--Finsler~spacetime} spacetime with the
leading symbol of a wide class of hyperbolic \index{hyperbolic~system}
partial differential equations.  In other parts of
parameter\index{parameter~space} space, the most natural
interpretation of the geometry is in terms of a bi-metric
\index{bi-metric~spacetime} spacetime, where one has a manifold that
is simultaneously equipped with two distinct pseudo-Riemannian metric
tensors. Further specialization in parameter\index{parameter~space}
space leads to a region where a single pseudo-Riemannian metric tensor
is encountered --- this mono-metric \index{mono-metric~spacetime}
regime corresponds to Lorentzian spacetimes of the type encountered in
standard general relativity\index{general~relativity} and
cosmology~\cite{Schutzhold:cosmic, Fischer, silke, frw}.  Thus
the analogue spacetime\index{analogue~spacetime} based on 2-component
BECs provides models not just for standard general relativistic
spacetimes, but also for the more general bi-metric
\index{bi-metric~spacetime}, and even more general pseudo--Finsler
\index{pseudo--Finsler~spacetime} spacetimes.

Additionally, the 2-BEC\index{2-BEC~system} system permits us to
provide a mass-generating \index{mass-generating~mechanism} mechanism
for the quasi-particle\index{quasi-particle}
excitations~\cite{mass1,mass2}.  The specific mass-generating
\index{mass-generating~mechanism} mechanism arising herein is rather
different from the Higgs mechanism of the standard model of particle
physics, and provides an interesting counterpoint to the more usual
ways that mass-generation\index{mass-generating~mechanism} is
achieved.  Furthermore, at short distances, where the ``quantum
\index{quantum~pressure} pressure'' term can no longer be neglected,
then even in the mono-metric \index{mono-metric~spacetime} regime one
begins to see deviations from ``Lorentz \index{Lorentz~invariance}
invariance'' --- and these deviations are qualitatively of the type
encountered in ``quantum gravity\index{quantum~gravity!phenomenology}
phenomenology'', with the interesting property that the Lorentz
\index{Lorentz~invariance!violation} violating physics is naturally
suppressed by powers of the quasi-particle\index{quasi-particle} mass
divided by the mass of the fundamental bosons that form the
condensate~\cite{qgp1,qgp2,qgp3,qgp4}. So in these analogue systems
the mass-generating \index{mass-generating~mechanism} mechanism is
related to the ``hierarchy problem'' and the suppression of
Lorentz-violating physics.  The 2-BEC model also allows us to probe
the ``universality'' (or lack thereof) in the Lorentz
\index{Lorentz~invariance!violation} violating
sector~\cite{qgp1,qgp2,qgp3,qgp4}. More generally, as one moves beyond
the hydrodynamic\index{hydrodynamic~limit} limit in generic
pseudo--Finsler\index{pseudo--Finsler~spacetime} parts of
parameter\index{parameter~space} space, one can begin to see hints of
geometrical structure even more general than the pseudo--Finsler
\index{pseudo--Finsler~spacetime} geometries.

While we do not wish to claim that the 2-BEC\index{2-BEC~system}
analogue spacetime of this chapter is necessarily a good model for the
real physical spacetime arising from the putative theory of ``quantum
gravity''\index{quantum~gravity} (be it string-model, loop-variable,
or lattice based), it is clear that the 2-BEC \index{2-BEC~system}
analogue spacetime\index{analogue~spacetime} is an extraordinarily
rich mathematical and physical structure that provides many
interesting hints regarding the sort of kinematics and dynamics that
one might encounter in a wide class of models for ``quantum
gravity\index{quantum~gravity!phenomenology} phenomenology''. This is
the fundamental reason for our interest in this model, and we hope we
can likewise interest the reader in this system and its relatives.

%%%%%%%%%%%%%%%%%%%%%%%%%%%%%%%%%%%%%%%%%%%%%%%%%%%

\section{Theory of the 2-component BEC}                          		               

%%%%%%%%%%%%%%%%%%%%%%%%%%%%%%%%%%%%%%%%%%%%%%%%%%%

The basis for our analogue model is an ultra-cold dilute atomic gas of
$N$ bosons, which exist in two single-particle states $\vert \A
\rangle$ and $\vert \B \rangle$.  For example, we consider two
different hyperfine\index{hyperfine~states} states, $\vert
F=1,m_{F}=-1 \rangle$ and $\vert F=2,m_{F}=2 \rangle$ of $^{87}\Rb$
\cite{jenkins,trippenbach}.  They have different total angular momenta
$F$ and therefore slightly different energies. That permits us, from a
theoretical point of view, to keep $m_{\A} \neq m_{\B}$, even if they
are very nearly equal (to about one part in $10^{16}$). At the assumed
ultra-cold temperatures and low densities the atoms interact only via
low-energy collisions, and the 2-body atomic potential can be replaced
by a contact potential. That leaves us with with three atom-atom
coupling constants, $U_{\A\A}$, $U_{\B\B}$, and $U_{\A\B}$, for the
interactions within and between the two
hyperfine\index{hyperfine~states} states. For our purposes it is
essential to include an additional laser field, that drives transition
between the two single-particle states.

%_figure__figure__figure__figure__figure__figure__figure__figure__figure__figure__figure__figure_
\begin{figure}[!htb]
 \begin{center}
 \input{energylevel.pstex_t}
 \caption[Energy levels for $\Rb^{87}$.]  {\label{energylevel}The
   horizontal lines indicate the hyperfine  states of $^{87}\Rb$. The
   arrows represent two laser fields --- with the two frequencies
   $\Omega_{1}$ and $\Omega_{2}$ --- necessary to drive transitions
   between the two trapped states $\vert F=1,m_{F}=-1 \rangle$ and
   $\vert F=2,m_{F}=2 \rangle$, where the frequency difference
   corresponds to the energy difference of the two hyperfine states.
   This is realized by a three-level atomic system, because the
   hyperfine states must be coupled over an intermediate level, that
   has to lie somewhat below the excited $\vert e \rangle$ states, as
   indicated by $\Delta$.}
 \end{center} 
 \index{hyperfine~states} 
 \end{figure}
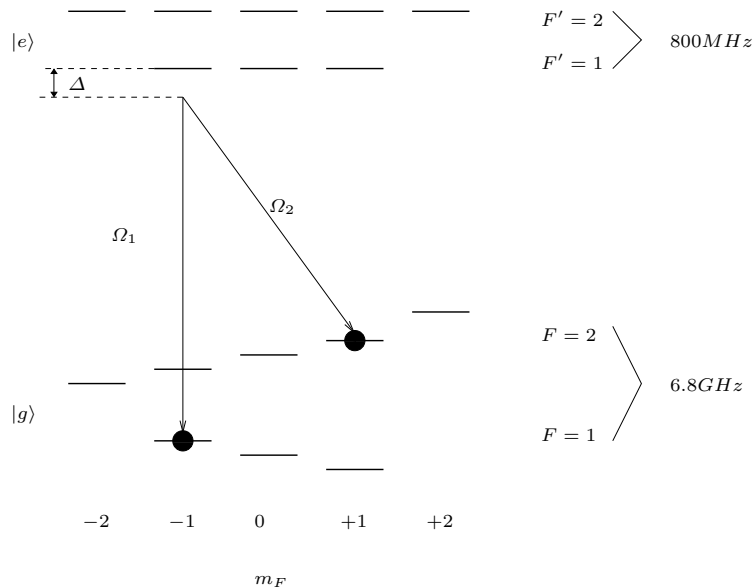
%_figure__figure__figure__figure__figure__figure__figure__figure__figure__figure__figure__figure_

 In Fig. \ref{energylevel} the energy levels for different
 hyperfine\index{hyperfine~states} states of $^{87}\Rb$, and possible
 transitions involving three-level processes are schematically
 explained. A more detailed description on how to set up an external
 field driving the required transitions can be found in \cite{Bloch}.

%+++++++++++++++++++++++++++++++++++++++++++++++++++++++++++++++++
\subsection{Gross--Pitaevskii equation}
%+++++++++++++++++++++++++++++++++++++++++++++++++++++++++++++++++

The rotating-frame Hamiltonian for our closed 2-component system is
given by:~\footnote{In general, it is possible that the collisions
  drive coupling to other hyperfine states. Strictly speaking the
  system is not closed, but it is legitimate to neglect this
  effect~\cite{dressed}. }\index{hyperfine~states}
\begin{eqnarray}
\hat H = \int \d \r \;
\Bigg\{ & \sum_{i = \A,\B} & \left(-\hat \Psi_{i}^\dag \frac{\hbar^2 \nabla^2}{2 m_{i}} \hat \Psi_{i}  
+ \hat \Psi_i^\dag V_{ext,i} (\r) \hat \Psi_i \right) 
\nonumber \\
+ \frac{1}{2}& \sum_ {i,j = \A,\B}& \left(
 U_{i j} \hat \Psi_i^\dag  \hat \Psi_j^\dag  \hat \Psi_i  \hat \Psi_j  
 + \lambda  \hat \Psi_i^\dag (\mathbf{\sigma}_{x})_{i j}  \hat \Psi_j
\right)
\Bigg\} \, ,
\end{eqnarray}
with the transition rate $\lambda$ between the two
hyperfine\index{hyperfine~states} states. Here $\hat \Psi_i(\r)$ and
$\hat \Psi_i^{\dag}(\r)$ are the usual boson field annihilation and
creation operators for a single-particle state at position $\r$, and
$\mathbf{\sigma}_x$ is the usual Pauli matrix.
For temperatures at or below the critical BEC temperature, almost all
atoms occupy the spatial modes $\Psi_\A(\r)$ and $\Psi_\B(\r)$. The
mean-field description for these modes,
\begin{equation}  \label{2GPE} 
 i \, \hbar \, \partial_{t} \Psi_{i} = \left[
   -\frac{\hbar^2}{2\,m_{i}} \nabla^2 + V_{i}-\mu_{i} + U_{ii}
   \, \betrag{\Psi_{i}}^2 + U_{ij} \betrag{\Psi_{j}}^2
   \right] \Psi_{i} + \lambda \, \Psi_{j} \, , 
\end{equation}
are a pair of coupled
Gross--Pitaevskii\index{Gross--Pitaevskii~equation} equations (GPE):
$(i,j)\rightarrow (\A,\B)$ or $(i,j)\rightarrow (\B,\A)$.

%+++++++++++++++++++++++++++++++++++++++++++++++++++++++++++++++++
\subsection{Dynamics}
%+++++++++++++++++++++++++++++++++++++++++++++++++++++++++++++++++

In order to use the above 2-component BEC as an analogue model, we
have to investigate small perturbations (sound waves) in the
condensate cloud.\footnote{The perturbations amplitude have to be
  small compared to the overall size of the condensate could, so that
  the system remains in equilibrium.}  The excitation spectrum is
obtained by linearizing around some background densities $\rho_{i0}$
and phases $\theta_{i0}$, using:
\begin{equation}
\Psi_{i}= \sqrt{\rho_{i0}+ \varepsilon \, \rho_{i1} }\,
e^{i(\theta_{i0}+ \varepsilon \, \theta_{i1} )}
\quad\hbox{for}\quad i=\A,\,\B \, .
\end{equation}
To keep the analysis as general as possible, we allow the two initial
background phases to be independent from each other, and define
\begin{equation}
\delta_{\A\B} \equiv \theta_{\A0} - \theta_{\B0},
\end{equation}
as their difference. \\

A tedious calculation \cite{mass1,mass2,qgp1} shows that it is
convenient to introduce the following $2 \times 2$ matrices: An
effective coupling matrix,
\begin{equation}
\hat{\Xi}=\Xi+\hat X, 
\end{equation}
where we introduced the energy-independent matrix
\begin{equation}
\Xi \equiv \frac{1}{\hbar}
\MatrixB{\tilde{U}_{\A\A}}{\tilde{U}_{\A\B}}{\tilde{U}_{\A\B}}{\tilde{U}_{\B\B}}.
\end{equation}
This matrix contains the quantities 
\begin{eqnarray}
\tilde{U}_{\A\A} &
%= U_{AA} -\frac{\lambda}{2}\frac{\sqrt{\rho_{B0}}}{\sqrt{\rho_{A0}}^{3}}
&\equiv U_{\A\A}-{\lambda \cos \delta_{\A\B} \; \sqrt{\rho_{\A0}\rho_{\B0}}\over2} {1\over \rho_{\A0}^2} , 
\\
\tilde{U}_{\B\B} &
%= U_{BB} -\frac{\lambda}{2}\frac{\sqrt{\rho_{A0}}}{\sqrt{\rho_{B0}}^{3}} 
&\equiv U_{\B\B}-{\lambda \cos \delta_{\A\B} \; \sqrt{\rho_{\A0}\rho_{\B0}}\over2} {1\over \rho_{\B0}^2}, 
\\
\tilde{U}_{\A\B} &
%= U_{AB}+ \frac{\lambda}{2}  \;\frac{1}{\sqrt{\rho_{A0} \, \rho_{B0} }}
&\equiv U_{\A\B}+{\lambda \cos \delta_{\A\B} \; \sqrt{\rho_{\A0}\rho_{\B0}}\over2} {1\over \rho_{\A0}\,\rho_{\B0}}.
\label{U_tilde_AB}
\end{eqnarray}
A second matrix, denoted $\hat X$, contains differential operators
$\hat Q_{X1}$ --- these are the second-order differential operators
obtained from linearizing the quantum
potential:\index{quantum~potential}
\begin{eqnarray}
 V_{\rm Q}(\rho_X) 
&\equiv&  
-  {\hbar^2\over2m_X} \left( {\nabla^2\sqrt{\rho_X}\over\sqrt{\rho_X}}\right) 
=
-  {\hbar^2\over2m_X} \left( {\nabla^2\sqrt{\rho_{X0}+\varepsilon\rho_{X1}}
\over\sqrt{\rho_{X0}+\varepsilon\rho_{X_1}} }\right) 
\\
&=& 
-  {\hbar^2\over2m_X} 
\left(  \hat Q_{X0}(\rho_{X0}) + \varepsilon \;\hat Q_{X1} (\rho_{X1}) \right).
\end{eqnarray}
The quantity $ \hat Q_{X0}(\rho_{X0})$ corresponds to the background
value of the quantum \index{quantum~pressure} pressure, and
contributes only to the background equations of motion --- it does not
affect the fluctuations. Now in a general background
\begin{equation}
\hat Q_{X1} (\rho_{X1})= {1\over2} \left\{ 
{ (\nabla\rho_{X0})^2-(\nabla^2\rho_{X0})\rho_{X0}\over\rho_{X0}^3} 
- {\nabla \rho_{X0}\over\rho_{X0}^2} \nabla 
+ {1\over\rho_{X0}} \nabla^2
\right\} \rho_{X1},
\end{equation}
and we define the matrix $\hat X$ to be
\begin{eqnarray}
\hat X &\equiv&
 -{\hbar\over2} \left[\begin{array}{cc} {\hat Q_{\A1}\over m_\A} & 0
    \\ 0 & {\hat Q_{\B1}\over m_\B}\end{array}\right].
\end{eqnarray}

Given the background homogeneity that will be appropriate for later
parts of the current discussion, this will ultimately simplify to
\begin{equation}
\hat Q_{X1} (\rho_{X1})=  {1\over2\rho_{X0}} \nabla^2 \rho_{X1},
\end{equation}
in which case
\begin{eqnarray}
\hat X &=&
-{\hbar\over4} \left[\begin{array}{cc} {1\over m_\A\;\rho_{\A0}} & 0
    \\ 0 & {1\over m_\B\;\rho_{\B0}} \end{array}\right]\nabla^2
=
 - X \; \nabla^2
\, .
\end{eqnarray}

Without transitions between the two hyperfine\index{hyperfine~states}
states, when $\lambda=0$, the matrix $\Xi$ only contains the coupling
constants $\Xi_{ij} \rightarrow U_{ij}/\hbar$. While $\Xi$ is
independent of the energy of the perturbations, the importance of
$\hat X$ increases with the energy of the perturbation. In the
so-called hydrodynamic \index{hydrodynamic~limit} approximation $\hat
X$ can be neglected, effectively $\hat X \rightarrow 0$ and $\hat \Xi
\rightarrow \Xi$.

Besides the interaction matrix, we also introduce a transition matrix,
\begin{equation}
\Lambda\equiv -\frac{2\lambda\, \cos \delta_{\A\B} \; \sqrt{\rho_{i0}\,\rho_{j0}} }{\hbar} \,
\MatrixB{+1}{-1}{-1}{+1}
\end{equation}
and a mass-density matrix,
\begin{equation} \label{m-d-matrix}
D \equiv \hbar \,  \MatrixB{\frac{ \rho_{\A0} } {m_{\A} }}{0}{0}{\frac{ \rho_{\B0} } {m_{\B} }}  
\equiv  \hbar \, \MatrixB{d_{\A}}{0}{0}{d_{\B}}.
\end{equation} 

The final step is to define two column vectors,  
\begin{equation}
{\bar{\theta}} \equiv [\theta_{\A1},\theta_{\B1}]^T,
\end{equation}
and  
\begin{equation}
{\bar{\rho}} \equiv [\rho_{\A1},\rho_{\B1}]^T.
\end{equation}
We then obtain two compact equations for the perturbation in the
phases and densities:
\begin{eqnarray} \label{thetavecdot}
\dot{\bar{\theta}}&=&  -\,\hat{\Xi} \; \bar{\rho} - \vec V  \cdot \nabla \bar{\theta} + \ipd \, \bar{\theta},
\\
\label{rhovecdot}
\dot{\bar{\rho}}&=& \, - \nabla \cdot \left( D \; \nabla \bar{\theta} +  \bar{\rho} \; \vec V \right) 
- \Lambda \;\bar{\theta} - \ipd^{\T} \bar{\rho} \, .
\end{eqnarray} 
Here the background velocity matrix simply contains the two
background velocities of each condensate,
\begin{equation}
\vec{V}= 
\left[
\begin{array}{cc}
{\vec{v}}_{\A0} & 0  \\
0 & {\vec{v}}_{\B0}  \\
\end{array}
\right] ,
\end{equation} 
with two possibly distinct background velocities,
\begin{equation}
\begin{split}
\vec{v}_{\A0} = & \frac{\hbar}{m_{\A}} \nabla \theta_{\A0}, \\
\vec{v}_{\B0} = & \frac{\hbar}{m_{\B}} \nabla \theta_{\B0} .\\
\end{split} 
\end{equation}
Additionally we also introduce the matrix $\ipd$, which depends on
the difference of the initial phases and is defined as
\begin{equation}
\ipd \equiv \frac{\lambda \sin \delta_{\A\B}}{\hbar} 
\left[
\begin{array}{cc}
+\sqrt{\frac{\rho_{\B0}}{ \rho_{\A0}}} & -\sqrt{\frac{\rho_{\B0}}{\rho_{\A0}}}  \\
+\sqrt{\frac{\rho_{\A0}}{\rho_{\B0}}} & -\sqrt{\frac{\rho_{\A0}}{\rho_{\B0}}} \\
\end{array}
\right].
\end{equation}

Now combine these two equations into one:
\begin{eqnarray} \label{Eq:GHWE} 
\partial_{t} (\hat{\Xi}^{-1} \; \dot{\bar{\theta}} ) =
& - &\partial_{t} \left(\hat{\Xi}^{-1} \; \vec V \cdot \nabla \bar{\theta} \right) 
 - \nabla (\vec V \; \hat{\Xi}^{-1} \; \dot{\bar{\theta}} ) \nonumber \\
& + &\nabla \cdot \left[ \left(D - \vec V \; \hat{\Xi}^{-1} \; \vec V \right) \nabla \bar{\theta} \, \right] 
 + \Lambda  \; \bar{\theta} \\ 
 &+& \K  \, \bar{\theta} + 
{1\over2}\left\{\G^{a} \partial_{a} \bar{\theta} + \partial_a (\G^a \bar{\theta} ) \right\} , 
\nonumber
\end{eqnarray}
where the index $a$ runs from $0$--$3$ (that is, over both time and
space), and we now define
\begin{eqnarray}
\G^{t} &=& \hat{\Xi}^{-1} \ipd - \ipd^{\T} \hat{\Xi}^{-1},  \\
\G^{i} &=& \vec V \hat{\Xi}^{-1} \ipd - \ipd^{\T} \hat{\Xi}^{-1} \vec V, 
\end{eqnarray}
and 
\begin{eqnarray}
\K &=& \ipd^{\T} \hat{\Xi}^{-1} \ipd + 
{1\over2} \partial_{t} \left( \hat{\Xi}^{-1} \ipd + \ipd^\T \hat{\Xi}^{-1} \right) + 
{1\over2} \nabla \left( \vec V \hat{\Xi}^{-1} \ipd + \ipd^\T \hat{\Xi}^{-1} \vec V \right).
\nonumber\\
&&
\end{eqnarray}
Note that the $\G^a$ matrices are antisymmetric in field-space ($\A \leftrightarrow \B$),
while the matrix $\K$ is symmetric. Also, both $\G^a\to 0$ and
$\K\to0$ as $\delta_{\A\B}\to0$.

Our first goal is to show that equation (\ref{Eq:GHWE}), which
fundamentally describes quasi-particle\index{quasi-particle}
excitations interacting with a condensed matter system in the
mean-field approximation, can be given a physical and mathematical
interpretation in terms of a classical background geometry for
massless and massive particles propagating through an analogue
spacetime~\cite{normal,normal2,birefringence,BEC}.\index{analogue~spacetime}
This analogy only holds (at least in its cleanest form) in the
so-called hydrodynamic \index{hydrodynamic~limit} limit
$\hat\Xi\to\Xi$, which limit is directly correlated with the
healing\index{healing~length} length which we shall now introduce.

%+++++++++++++++++++++++++++++++++++++++++++++++++++++++++++++++++
\subsection{Healing length\label{sec:heal}}
%+++++++++++++++++++++++++++++++++++++++++++++++++++++++++++++++++

The differential operator $\hat Q_{X1}$ that underlies the origin of
the $\hat X$ contribution above is obtained by linearizing the
quantum\index{quantum~potential} potential
\begin{equation}
 V_{\rm Q}(\rho_X) \equiv  -  {\hbar^2\over2 m_X}
 \left( {\nabla^2 \sqrt{\rho_X}\over\sqrt{\rho_X}}\right)
\end{equation}
which appears in the Hamilton--Jacobi\index{Hamilton--Jacobi~equation}
equation of the BEC flow.  This quantum
potential\index{quantum~potential} term is suppressed by the smallness
of $\hbar$, the comparative largeness of $m_X$, and for sufficiently
uniform density profiles.  But of course in any real system the
density of a BEC must go to zero at the boundaries of its electro-magnetic trap
(given that $\rho_X=|\psi_X(\vec x,t)|^2$).  In a 1-component BEC the
healing\index{healing~length} length characterizes the minimal
distance over which the order parameter goes from zero to its bulk
value. If the condensate density grows from zero to $\rho_0$ within a
distance $\xi$ the quantum\index{quantum~potential} potential term
(non local) and the interaction energy (local) are respectively
$E_{\rm kinetic}\sim \hbar^2/(2m\xi^2)$ and $E_{\rm interaction}\sim
4\pi\hbar^2 a \rho_0/m$. These two terms are comparable when
 \begin{equation}
\xi=(8\pi \rho_0 a)^{-1/2},
  \label{heal}
 \end{equation}
 where $a$ is the $s$-wave scattering\index{scattering~length} length
 defined as
\begin{equation}
a = {m \; U_0\over4\pi \hbar^2}.
\end{equation}
Note that what we call $U_0$ in the above expression is just the
coefficient of the non-linear self-coupling term in the
Gross--Pitaevskii\index{Gross--Pitaevskii~equation} equation, \ie,
just $U_{\A\A}$ or $U_{\B\B}$ if we completely decouple the 2 BECs
($U_{\A\B}=\lambda=0$).

Only for excitations with wavelengths much larger than the
healing\index{healing~length} length is the effect of the quantum
potential\index{quantum~potential} negligible. This is called the
hydrodynamic \index{hydrodynamic~limit} limit because the
single--BEC\index{Bose--Einstein~condensate} dynamics is then
described by the continuity and Hamilton--Jacobi equations of a
super-fluid, and its excitations behave like massless phononic modes.
In the case of excitations with wavelengths comparable with the
healing\index{healing~length} length this approximation is no longer
appropriate and deviations from phononic behaviour will arise.

Such a simple discrimination between different regimes is lost once
one considers a system formed by two coupled
Bose--Einstein\index{Bose--Einstein~condensate} condensates. One is
forced to introduce a generalization of the healing $\xi$
length\index{healing~length} in the form of a ``healing matrix''. If
we apply the same reasoning used above for the definition of the
``healing\index{healing~length} length'' to the 2-component
BEC\index{2-BEC~system} system we again find a functional form like
that of equation~(\ref{heal}) however we now have the crucial
difference that both the density and the
scattering\index{scattering~length} length are replaced by matrices.
In particular, we generalize the scattering\index{scattering~length}
length $a$ to the matrix $\mathcal{A}$:
 \begin{equation}
 \mathcal{A} = {1\over4\pi\hbar^2} 
 \left[\begin{array}{cc}\sqrt{m_\A}&0\\0&\sqrt{m_\B}\end{array}\right]
 \;
 \left[\begin{array}{cc}
 \tilde U_{\A\A}&\tilde U_{\A\B}\\ \tilde U_{\A\B} &\tilde U_{\B\B} 
 \end{array}\right]
 \;
 \left[\begin{array}{cc}\sqrt{m_\A}&0\\0&\sqrt{m_\B}\end{array}\right].
 \end{equation}
 Furthermore, from (\ref{heal}) a healing\index{healing~length} length
 matrix $Y$ can be defined by
 \begin{equation}
Y^{-2} = {2\over\hbar^2} \left[\begin{array}{cc}
\sqrt{\rho_{\A0} m_\A}&0\\0&\sqrt{\rho_{\B0} m_\B}\end{array}\right]
 \;
 \left[\begin{array}{cc}
 \tilde U_{\A\A}&\tilde U_{\A\B}\\ \tilde U_{\A\B} &\tilde U_{\B\B} 
 \end{array}\right]
 \;
 \left[\begin{array}{cc}
 \sqrt{\rho_{\A0} m_\A}&0\\0&\sqrt{\rho_{\B0} m_\B}\end{array}\right].
 \end{equation}
 That is, in terms of the matrices we have so far defined:
 \begin{equation}
 Y^{-2} = {1\over2} \;X^{-1/2} \; \Xi \; X^{-1/2};
 \qquad
 Y^2 = 2\; X^{1/2} \; \Xi^{-1} \; X^{1/2}.
 \end{equation}
 Define ``effective'' scattering\index{scattering~length} lengths and
 healing\index{healing~length} lengths for the 2-BEC
 \index{2-BEC~system} system as
\begin{equation}
a_\mathrm{eff} = {1\over2}\;\tr[{\mathcal{A}}] =
 {m_\A \tilde U_{\A\A} + m_\B \tilde U_{\B\B}\over 8\pi\hbar^2},
\end{equation}
and
\begin{equation}
\xi_\mathrm{eff}^2 = {1\over2}\;\tr[Y^2] = \tr[X \Xi^{-1}] 
= {\hbar^2 [\tilde U_{\B\B}/(m_\A \rho_{\A0}) + \tilde U_{\A\A}/(m_\B \rho_{\B0} )]\over
4 (\tilde U_{\A\A} \tilde U_{\B\B} - \tilde U_{\A\B}^2 ) }.
\end{equation}
That is
\begin{equation}
\xi_\mathrm{eff}^2 = 
{\hbar^2 [m_\A \rho_{\A0} \tilde U_{\A\A}+ m_\B \rho_{\B0} \tilde U_{\B\B}]
\over
4 m_\A m_\B \rho_{\A0} \rho_{\B0} \;(\tilde U_{\A\A} \tilde U_{\B\B} - \tilde U_{\A\B}^2 ) }.
\end{equation}
Note that if the two components are decoupled and tuned to be
equivalent to each other, then these effective
scattering\index{scattering~length} and healing\index{healing~length}
lengths reduce to the standard one-component results.

%%%%%%%%%%%%%%%%%%%%%%%%%%%%%%%%%%%%%%%%%%%%%%%%%%%

\section{Emergent spacetime at low energies}
             		                            
%%%%%%%%%%%%%%%%%%%%%%%%%%%%%%%%%%%%%%%%%%%%%%%%%%%

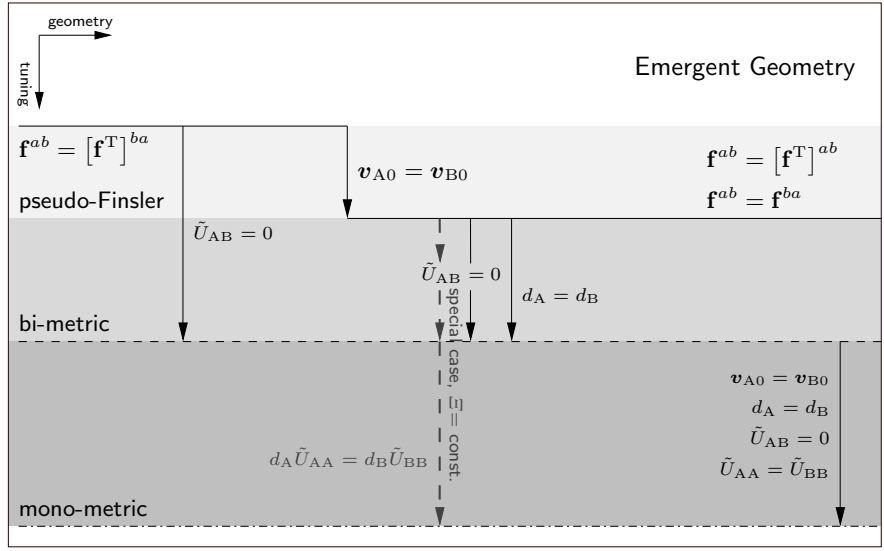
\begin{figure}[!htb]
 \begin{center}
 \input{geometry.pstex_t}
\caption[Concept for fine tuning.]
{\label{Geometry}The dependence of the emergent geometry on the 2-component BEC parameters.}       
 \end{center}         
\end{figure}

The basic idea behind analogue models is to re-cast the equation for
excitations in a fluid into the equation describing a massless or
massive scalar field embedded in a pseudo--Riemannian geometry.
Starting from a two component superfluid we are going to show that it
is not only possible to obtain a massive scalar field from such an
analogue model, in addition we are also able to model much more
complex geometries.  In Fig. \ref{Geometry} we illustrate how
excitations in a 2-component BEC are associated with various types of
emergent \index{emergent~geometry} geometry.

%_figure__figure__figure__figure__figure__figure__figure__figure__figure__figure__figure__figure_
%_figure__figure__figure__figure__figure__figure__figure__figure__figure__figure__figure__figure_

Most generally, we show that excitations in a 2-component BEC (in the
hydrodynamic \index{hydrodynamic~limit} limit) can be viewed as
propagating through a specific class of pseudo--Finsler
\index{pseudo--Finsler~spacetime} geometry.
As additional constraints are placed on the BEC
parameter\index{parameter~space} space, the geometry changes from
pseudo--Finsler \index{pseudo--Finsler~spacetime}, first to bi-metric,
and finally to mono-metric \index{mono-metric~spacetime}
(pseudo--Riemannian, Lorentzian) geometry.
This can be accomplished by tuning the various BEC parameters, such as
the transition rate $\lambda$, the background velocities $\vec
v_{\A0}$, $\vec v_{\B0}$, the background densities $\rho_{\A0}$,
$\rho_{\B0}$, and the coupling between the atoms $U_{\A \A}$, $U_{\B
  \B}$ and $U_{\A\B}$.

At first, it might seem to be quite an artificial thing to impose such
constraints onto the system. But if one considers that the two
macroscopic wave functions represent two interacting classical fields,
it is more or less obvious that this is the \emph{only} way in which
to enforce physical constraints onto the fields themselves, and on the
way they communicate with each other.
%

%+++++++++++++++++++++++++++++++++++++++++++++++++++++++++++++++++
\subsection{Pseudo-Finsler geometry \label{GHWE}}
%+++++++++++++++++++++++++++++++++++++++++++++++++++++++++++++++++

In the hydrodynamic \index{hydrodynamic~limit} limit ($\hat{\Xi} \rightarrow \Xi$), it is
possible to simplify equation (\ref{Eq:GHWE}) --- without enforcing
any constraints on the BEC parameters --- if we adopt a
(3+1)-dimensional ``spacetime'' notation by writing $x^a=(t,x^i)$, with
$i\in\{1,2,3\}$ and $a\in\{0,1,2,3\}$. Then 
equation  (\ref{Eq:GHWE}) can be very compactly rewritten as~\cite{normal,normal2}:
\begin{equation}
\label{E:fab_general}
\partial_a \left( \mathbf{f}^{ab} \; \partial_b  \bar\theta \right) 
+ \left(\Lambda +  \K \right)  \; \bar{\theta} 
+ {1\over2}\left\{ 
\G^{a} \; \partial_{a} {\bar{\theta}}  + \partial_a (\G^a \bar\theta) 
\right\}= 0.
\end{equation}
The object $\mathbf{f}^{ab}$ is a $4\times 4$ spacetime matrix
(actually a tensor density), each of whose components is a $2\times 2$
matrix in field-space --- equivalently this can be viewed as a
$2\times 2$ matrix in field-space each of whose components is a
$4\times 4$ spacetime tensor density. By inspection this is a
self-adjoint second-order linear system of PDEs. The spacetime
geometry is encoded in the leading-symbol of the PDEs, namely the
$\mathbf{f}^{a b}$, without considering the other subdominant
terms. That this is a sensible point of view is most easily seem by
considering the usual curved-spacetime d'Alembertian equation for a
charged particle interacting with a scalar potential in a standard
pseudo--Riemannian geometry
\begin{equation}
{1\over\sqrt{-g}} [\partial_a - i A_a] 
\left( \phantom{\Big{|}} \sqrt{-g} g^{ab} [\partial_b-iA_b] \theta \right) + V \theta = 0
\end{equation}
from which it is clear that we want to make the analogy
\begin{equation}
\mathbf{f}^{ab} \sim  \sqrt{-g} \; g^{ab}
\end{equation}
as the key quantity specifying the geometry. In addition
\begin{equation}
\G^a \sim i A^a \qquad \hbox{and} \qquad \Lambda + \K \sim V - g^{ab} A_a A_b
\end{equation}
so that $\G^a$ is analogous to a vector potential and $\Lambda$ (plus
corrections) is related to the scalar potential $V$ --- in a
translation invariant background this will ultimately provide a mass
term.

Specifically in the current 2-BEC \index{2-BEC~system} system we have
\begin{equation} \label{Finsler_1}
\mathbf{f}^{a b}=
\left(
\begin{array}{c|c}
-\Xi^{-1}            & - ( \vec{V} \Xi^{-1} )^{\T} \\
\hline
%--- & & ---------- \\
\vphantom{\Big|}
- \vec{V} \Xi^{-1}  & 
D - \vec{V}\Xi^{-1} \vec{V}^{\,T}
\end{array}
\right),
\end{equation}
where
\begin{equation}
\vec{V}^{\,\T} = \left[
\begin{array}{cc}
\vec{v}_{\A0}^{ \T} & 0 \\ 0 & \vec{v}_{\B0}^{\T}
\end{array} \right]
\end{equation}
is a $2\times2$ matrix in field space that is also a row vector in
physical 3-space.  Overall, this does look like a rather complicated
object. However, it is possible to re-write the $4\times4$ geometry
containing $2\times 2$ matrices as its elements, in form of a single
$(2\cdot4) \times (2\cdot 4)$ matrix.\footnote{This result can be
  generalized for $n$-component systems. Any $4 \times 4$ geometry
  obtained from a $n$-component system can be re-written as a single
  $(n\cdot4) \times (n\cdot 4)$ matrix.}  Explicitly we
have\footnote{Note $\Xi_{12}= \Xi_{21}$, so $\Xi_{12}^{-1}=
  \Xi_{21}^{-1}$.}
\begin{eqnarray} \nonumber
&&
\mathbf{f}^{a b}=
\left[ \begin{array}{c|c}
{\Xi}^{-1}_{11} 
\left(\begin{array}{c|c} -1  & - \vec{v}_{\A0}^{\T} \\ \hline
	- \vec{v}_{\A0} & \frac{ d_{\A}}{{\Xi}^{-1}_{11} }\delta_{ij} - \vec{v}_{\A0} \vec{v}_{\A0}^{\T}
	\end{array} \right)
& 
{\Xi}^{-1}_{12} 
\left(\begin{array}{c|c} 1  &  \vec{v}_{\B0}^{\T} \\ \hline
	\vec{v}_{\A0} &  \vec{v}_{\A0} \vec{v}_{\B0}^{\T}
	\end{array} \right)
\\ \hline
{\Xi}^{-1}_{21} 
\left( \begin{array}{c|c} 1  &  \vec{v}_{\A0}^{\T} \\ \hline \vec{v}_{\B0} &  \vec{v}_{\B0} \vec{v}_{\A0}^{\T}
	\end{array} \right)
& 
{\Xi}^{-1}_{22} 
\left( \begin{array}{c|c} -1  & - \vec{v}_{\B0}^{\T} \\ \hline - \vec{v}_{\B0} & \frac{d_{\B} }{{\Xi}^{-1}_{22}}\delta_{ij} - \vec{v}_{\B0} \vec{v}_{\B0}^{\T} \end{array}
\right) \end{array} \right],
\\
&&
\end{eqnarray}
which we can re-write as
\begin{equation} 
\mathbf{f}=
\left[
\begin{array}{c|c}
-{\Xi}^{-1}_{11} \, \mathcal{V}_{1}\mathcal{V}_{1}^{\T} + D_{11} h
& 
-{\Xi}^{-1}_{12} \, \mathcal{V}_{1}\mathcal{V}_{2}^{\T}+ D_{12} h
\\
\hline
-{\Xi}^{-1}_{21} \, \mathcal{V}_{2}\mathcal{V}_{1}^{\T}+ D_{21} h
& 
-{\Xi}^{-1}_{22} \, \mathcal{V}_{2}\mathcal{V}_{2}^{\T}+ D_{22} h
\end{array}
\right],
\end{equation}
where
\begin{eqnarray}
\mathcal{V}_{1}^a &:=& \left(1,\vec{v}_{\A0}^i \right), \\
\mathcal{V}_{2}^a &:=& \left(1,\vec{v}_{\B0}^i \right),
\end{eqnarray}
and
\begin{equation}
h^{ab}:= \diag(0,1,1,1).
\end{equation}

Even simpler is the form
\begin{equation} 
\mathbf{f}=
\left[
\begin{array}{c|c}
-{\Xi}^{-1}_{11} \, \mathcal{V}_{1}\mathcal{V}_{1}^{\T} 
& 
-{\Xi}^{-1}_{12} \, \mathcal{V}_{1}\mathcal{V}_{2}^{\T}
\\
\hline
-{\Xi}^{-1}_{21} \, \mathcal{V}_{2}\mathcal{V}_{1}^{\T} 
& 
-{\Xi}^{-1}_{22} \, \mathcal{V}_{2}\mathcal{V}_{2}^{\T} 
\end{array}
\right]
+
D \otimes h.
\end{equation}
The key point is that this allows us to write
\begin{equation} 
\label{pseudo--Finsler-metric}
\mathbf{f}^{a b}=
\MatrixB{f_{11}^{a b}}{f_{12}^{a b}}{f_{21}^{a b}}{f_{22}^{a b}},
\end{equation}
where
\begin{equation}
\begin{split}
f_{11}^{a b}=& 
-{\Xi}^{-1}_{11} \, \mathcal{V}_{1}^{a}\mathcal{V}_{1}^{b} + D_{11} h^{a b}, \\
f_{12}^{a b}=&
-{\Xi}^{-1}_{12} \, \mathcal{V}_{1}^{a}\mathcal{V}_{2}^{b}, \\
f_{21}^{a b}=&
-{\Xi}^{-1}_{12} \, \mathcal{V}_{2}^{a}\mathcal{V}_{1}^{b}, \\
f_{22}^{a b}=&
-{\Xi}^{-1}_{22} \, \mathcal{V}_{2}^{a}\mathcal{V}_{2}^{b} + D_{22} h^{a b}. \\
\end{split}
\end{equation}

It is also possible to separate the representation of
$\mathbf{f}^{ab}$ into field space and position space as follows
\begin{equation} 
\label{Finsler_product}
\begin{split}
\mathbf{f}^{a b} =&
\MatrixB{{\Xi}^{-1}_{11}}{0}{0}{0}  \mathcal{V}_{1}^{a} {\mathcal{V}_{1}^{b}}
+\MatrixB{0}{0}{0}{{\Xi}^{-1}_{22}} \mathcal{V}_{2}^{a} {\mathcal{V}_{2}^{b}} \\
&+\MatrixB{0}{{\Xi}^{-1}_{12}}{0}{0}  \mathcal{V}_{1}^{a} {\mathcal{V}_{2}^{b}} +
\MatrixB{0}{0}{{\Xi}^{-1}_{21}}{0} \mathcal{V}_{2}^{a} {\mathcal{V}_{1}^{b}} +
 D  h^{a b}.
\end{split}
\end{equation}

Why do we assert that the quantity $\mathbf{f}^{ab}$ defines a
pseudo--Finsler \index{pseudo--Finsler~spacetime} geometry? (Rather
than, say, simply a $2\times2$ matrix of ordinary Lorentzian
geometries?) To see the reason for this claim, recall the standard
result~\cite{Courant} that the leading symbol of a system of PDEs
determines the ``signal speed'' (equivalently, the characteristics, or
the causal structure)~\cite{normal2}. Indeed if we consider the
eikonal\index{eikonal} approximation (while still remaining in the
realm of validity of the hydrodynamic \index{hydrodynamic~limit}
approximation) then the causal structure is completely determined by
the leading term in the Fresnel\index{Fresnel~equation} equation
\begin{equation}
\det[ \mathbf{f}^{ab} k_a k_b] = 0,
\end{equation}
where the determinant is taken in field space. (The quantity
$\mathbf{f}^{ab} k_a k_b$ is exactly what is called the leading symbol
of the system of PDEs, and the vanishing of this determinant is the
statement that high-frequency modes can propagate with wave vector
$k_a$, thereby determining both characteristics and causal structure.)
In the 2-BEC \index{2-BEC~system} case we can explicitly expand the
determinant condition as
\begin{equation}
(f_{11}^{ab} k_a k_b) (f_{22}^{cd} k_c k_d) 
-
(f_{12}^{ab} k_a k_b) (f_{21}^{cd} k_c k_d) = 0.
\end{equation}
Define a completely symmetric rank four tensor
\begin{equation}
Q^{abcd} \equiv f_{11}^{(ab} f_{22}^{cd)}
-
f_{12}^{(ab} f_{21}^{cd)},
\end{equation}
then the determinant condition is equivalent to
\begin{equation}
Q^{abcd}  k_a k_b k_c k_d = 0,
\end{equation}
which now defines the characteristics in terms of the vanishing of the
pseudo-co--Finsler \index{co--Finsler~geometry} structure 
\begin{equation}
Q(k) = Q^{abcd}  k_a k_b k_c k_c,
\end{equation}
defined on the cotangent bundle. As explained in appendix A, this
pseudo-co--Finsler \index{co--Finsler~geometry} structure can be
Legendre transformed to provide a pseudo--Finsler
\index{pseudo--Finsler~spacetime} structure, a Finslerian notion of
distance
\begin{equation}
ds^4 = g_{abcd} \; dx^a dx^b dx^c dx^d.
\end{equation}
Here the completely symmetric rank 4 tensor $g_{abcd}$ determines the
``sound cones'' through the relation $ds=0$. It is interesting to note
that a distance function of the form
\begin{equation}
ds = \sqrt[4]{g_{abcd} \; dx^a dx^b dx^c dx^d}
\end{equation}
first made its appearance in Riemann's inaugural lecture of
1854~\cite{Riemann}, though he did nothing further with it, leaving it
to Finsler \index{Finsler!geometry} to develop the branch of geometry
now bearing his name~\cite{Finsler}. The present discussion is
sufficient to justify the use of the term ``pseudo--Finsler'' in the
generic 2-BEC \index{2-BEC~system} situation, but we invite the more
mathematically inclined reader to see appendix A for a sketch of how
much further these ideas can be taken.

The pseudo--Finsler \index{pseudo--Finsler~spacetime} geometry
implicit in (\ref{pseudo--Finsler-metric}) is rather complicated
compared with the pseudo-Riemannian geometry we actually appear to be
living in, at least as long as one accepts standard general relativity
\index{general~relativity} as a good description of reality.
To mimic real gravity, we need to simplify our model.
It is now time to use the major advantage of our analogue model, the
ability to tune the BEC parameters, and with it the 2-field background
configuration.  The first order of business is to decouple
$\mathbf{f}^{ab}$ in field space.

%+++++++++++++++++++++++++++++++++++++++++++++++++++++++++++++++++
\subsection{Bi-metric geometry \label{bi-metric}}
%+++++++++++++++++++++++++++++++++++++++++++++++++++++++++++++++++

The reduction of equation (\ref{Finsler_product}) to a diagonal
representation in field space (via an orthogonal rotation on the
fields),
\begin{equation} \label{f_diagonal}
\mathbf{f}^{ab} \to \diag \left[ f^{ab}_{11},f^ {ab}_{22}\right] 
= \diag \left[ \sqrt{-g_{11}}\, g^{ab}_{11},\sqrt{-g_{22}}\, g^{ab}_{22} \right],
\end{equation}
enforces a bi-metric \index{bi-metric~spacetime} structure onto the
condensate.  There are two ways to proceed.
%~~~~~~~~~~~~~~~~~~~~~~~~~~~~~~~~~~~~~~~~~~~~~~~~~~~~
\subsubsection{Distinct background velocities}
%~~~~~~~~~~~~~~~~~~~~~~~~~~~~~~~~~~~~~~~~~~~~~~~~~~~~
For
\begin{equation}
\mathcal{V}_{1} \neq \mathcal{V}_{2},
\end{equation}
we require all five $2\times2$ matrices appearing in
(\ref{Finsler_product}) to commute with each other. This has the
unique solution $\Xi^{-1}_{12}=0$, whence
\begin{equation}
\tilde{U}_{\A\B} = 0.
\end{equation}
%%%
We then get
\begin{equation} 
\label{bi-metric_product}
\mathbf{f}^{a b} =
\MatrixB{{\Xi}^{-1}_{11}}{0}{0}{0}  \mathcal{V}_{1}^{a} {\mathcal{V}_{1}^{b}}
+\MatrixB{0}{0}{0}{{\Xi}^{-1}_{22}} \mathcal{V}_{2}^{a} {\mathcal{V}_{2}^{b}} + D  h^{a b}.
\end{equation}
Since $D$ is a diagonal matrix this clearly represents a bi-metric
\index{bi-metric~spacetime} geometry.  The relevant parameters are
summarized in Table \ref{bi-metricity_results}.

%~~~~~~~~~~~~~~~~~~~~~~~~~~~~~~~~~~~~~~~~~~~~~~~~~~~~
\subsubsection{Equal background velocities}
%~~~~~~~~~~~~~~~~~~~~~~~~~~~~~~~~~~~~~~~~~~~~~~~~~~~~
For
\begin{equation}
\mathcal{V}_{1} =\mathcal{V}_{2} \equiv \mathcal{V},
\end{equation}
we are still dealing with a pseudo--Finsler
\index{pseudo--Finsler~spacetime} geometry, one which is now
independently symmetric in field space
($\mathbf{f}^{ab}=[\mathbf{f}^\T]^{ab}$), and position space
$\mathbf{f}^{ab}=\mathbf{f}^{ba}$.\footnote{The most general
  pseudo--Finsler geometry is symmetric under simultaneous exchange of
  field space and position space:
  $\mathbf{f}^{ab}=[\mathbf{f}^\T]^{ba}$.}\index{pseudo--Finsler~spacetime}
In terms of the BEC parameters that means we must set equal the two
background velocities, $\vec{v}_{\A0}=\vec{v}_{\B0}\equiv\vec{v}_{0}$,
and equation (\ref{Finsler_product}) is simplified to:
\begin{equation} \label{Finsler_tuned}
\mathbf{f}^{ab} = - \Xi^{-1} \mathcal{V}^a \mathcal{V}^b + D h^{ab}.
\end{equation}
{From} the above, diagonalizability in field space now additionally
requires the commutator of the interaction and mass-density matrix to
vanish:
\begin{equation}
[\Xi,D]=0 \quad \implies \quad \tilde{U}_{\A\B}(d_{\A}-d_{\B})=0.
\end{equation}
Here, we have a choice between two tuning conditions that do the job:
\begin{equation}
\tilde{U}_{\A\B}=0 \qquad \hbox{or} \qquad d_{\A}=d_{\B}.
\end{equation}

Under the first option, where $\,\tilde{U}_{\A\B}=0$, the two
off-diagonal elements in equation (\ref{Finsler_tuned}) are simply
zero, and we get the desired bi-metricity in the form\footnote{We
  would like to stress that this constraint can be easily fulfilled,
  at least in the special case $\delta_{\A\B}=0$, by tuning the
  transition rate $\lambda$, see equation (\ref{U_tilde_AB}).}
\begin{equation} 
\label{bi-metric_product2}
\mathbf{f}^{a b} =
\MatrixB{{\Xi}^{-1}_{11}}{0}{0}{{\Xi}^{-1}_{22}}  \mathcal{V}^{a} {\mathcal{V}^{b}}
+ D  h^{a b}.
\end{equation}

Under the second option, for $d_{\A}=d_{\B}\equiv d$, we have $D =
d\,\mathbf{I}$. The situation is now a bit trickier, in the sense
that one has to diagonalize ${\Xi}^{-1}$:
\begin{equation}
\begin{split}
\tilde{\Xi}^{-1} &= O^{\T} \, \Xi^{-1} O\\ &= \diag
\scriptstyle{
\left[
\frac{\tilde{U}_{\A\A}+\tilde{U}_{\B\B}+\sqrt{(\tilde{U}_{\A\A}-\tilde {U}_{\B\B})^{2}+4\tilde{U}_{\A\B}^{2}}}
       {2\,(\tilde{U}_{\A\A}\tilde{U}_{\B\B}-\tilde{U}_{\A\B}^{2})} ,
\frac{\tilde{U}_{\A\A}+\tilde{U}_{\B\B}-\sqrt{(\tilde{U}_{\A\A}-\tilde {U}_{\B\B})^{2}+4\tilde{U}_{\A\B}^{2}}}
      {2\,(\tilde{U}_{\A\A}\tilde{U}_{\B\B}-\tilde{U}_{\A\B}^{2})}
\right] }.
\end{split}
\end{equation}
Once this is done, the way to proceed is to use the elements of
$\tilde{\Xi}^{-1}$ instead of ${\Xi}^{-1}$ in equation (\ref
{bi-metric_product2}).
The relevant parameters are summarized in Table \ref{bi-metricity_results}.

\begin{table}[htdp]
\begin{center}
\tiny
\begin{tabular}{|c||c|c|c|}
\hline
          \multicolumn{4}{|c|}{\rule[-3mm]{0mm}{8mm}\small \textsf{Bi-metric tuning  scenarios}}   \\
\hline
      &\rule[-2mm]{0mm}{6mm} \small $\vec{v}_{\A0}\neq \vec{v}_{\B0}$  &  \multicolumn{2}{|c|}{\small $ \vec{v}_{\A0}=\vec{v}_{\B0}$}  \\
\hline
     &\rule[-2mm]{0mm}{6mm}  \small $\tilde{U}_{\A\B}=0$ & \small $\tilde{U}_{\A\B}=0$ & \small $d_{\A} =d_{\B}$  \\
\hline
\hline
& & &  \\
\small $f_{11}^{ab}\propto$
&
$ \left( \begin{array}{c|c}
- 1 & - \vec{v}_{\A0}^{\T} \\ \hline
-\vec{v}_{\A 0} & \tilde{U}_{\A\A}  d_{\A} \, h^{ij} -  \vec{v}_{\A0} \vec{v}_{\A0}^{\T} 
\end{array} \right)$
&
$ \left( \begin{array}{c|c}
- 1 & - \vec{v}_{0}^{\T} \\ \hline
-\vec{v}_{0} & \tilde{U}_{\A\A}  d_{\A} \, h^{ij} -  \vec{v}_{0}\vec{v} _{0}^{\T}
\end{array} \right)$
&
$ \left( \begin{array}{c|c}
- 1 & - \vec{v}_{0}^{\T} \\ \hline
-\vec{v}_{0} & \tilde{\Xi}^{-1}_{11}  d \, h^{ij} -  \vec{v}_{0}\vec {v}_{0}^{\T}
\end{array} \right)$\\
& & &  \\
\small $f_{22}^{ab} \propto$ &
$ \left( \begin{array}{c|c}
- 1 & - \vec{v}_{\B0}^{\T} \\ \hline
-\vec{v}_{\B 0} & \tilde{U}_{\B\B}  d_{\B} \, h^{ij} -  \vec{v}_{\B0} \vec{v}_{\B0}^{\T} 
\end{array} \right)$
&
$ \left( \begin{array}{c|c}
- 1 & - \vec{v}_{0}^{\T} \\ \hline
-\vec{v}_{0} & \tilde{U}_{\B\B}  d_{\B} \, h^{ij} -  \vec{v}_{0}\vec{v} _{0}^{\T}
\end{array} \right)$
&
$ \left( \begin{array}{c|c}
- 1 & - \vec{v}_{0}^{\T} \\ \hline
-\vec{v}_{0} & \tilde{\Xi}^{-1}_{22}  d \, h^{ij} -  \vec{v}_{0}\vec {v}_{0}^{\T}
\end{array} \right)$\\
& & &  \\
\hline
& & \multicolumn{2}{|c|}{}  \\
\small ${g_{11}}_{ab}\propto$  &
\small $\left( \begin{array}{c|c}
-(c_{11}^{2} - v_{\A0}^{2}) & - \vec{v}_{\A0}^{\T} \\ \hline
-\vec{v}_{\A0} &  h^{ij} \end{array} \right) $&
\multicolumn{2}{|c|}{
\small $\left( \begin{array}{c|c}
-(c_{11}^{2} - v_{0}^{2}) & - \vec{v}_{0}^{\T} \\ \hline
-\vec{v}_{0} &  h^{ij} \end{array} \right) $}
\\
& & \multicolumn{2}{|c|}{}  \\
\small ${g_{22}}_{ab}\propto$ &
\small $\left( \begin{array}{c|c}
-(c_{22}^{2} - v_{\B0}^{2}) & - \vec{v}_{\B0}^{\T} \\ \hline
-\vec{v}_{\B0} &  h^{ij} \end{array} \right) $&
\multicolumn{2}{|c|}{
\small $\left( \begin{array}{c|c}
-(c_{22}^{2} - v_{0}^{2}) & - \vec{v}_{0}^{\T} \\ \hline
-\vec{v}_{0} &  h^{ij} \end{array} \right) $}
\\
& & \multicolumn{2}{|c|}{}  \\
\hline
& \multicolumn{2}{|c|}{ } &  \\
\small $c_{11}^{2}=$ &
\multicolumn{2}{|c|}{ \small
$\tilde{U}_{\A\A} d_{\A} = \frac{ U_{\A\A} \rho_{\A0} + U_{\A\B}\rho_ {\B0}}{m_{\A}}$}&
\small $\tilde{\Xi}^{-1}_{11} \, d$
\\
& \multicolumn{2}{|c|}{ } &  \\
\small $c_{22}^{2}=$ &
\multicolumn{2}{|c|}{ \small
$\tilde{U}_{\B\B} d_{\B} = \frac{ U_{\B\B} \rho_{\B0} + U_{\A\B}\rho_ {\A0}}{m_{\B}}$}&
\small $\tilde{\Xi}^{-1}_{22} \, d $
\\
& \multicolumn{2}{|c|}{ } &  \\
\hline
\end{tabular}
\caption{\label{bi-metricity_results} If the pseudo--Finsler geometry
  decouples into two independent Lorentzian geometries $f^{ab}_{11} =
  \sqrt{-g_{11}} g_{11}$ and $f^{ab}_{11} = \sqrt{-g_{11}} g_{11}$,
  with two distinct speed of sounds $c_{11}$ and $c_{22}$, we are
  effectively dealing with a bi-metric Lorentzian metric. The table
  shows the results from three different tuning scenarios, that are
  sufficient to drive the 2-component BEC from Finsler to
  bi-Lorentzian spacetime. The rightmost column $d_A=d_B$ is addressed in \cite{Fischer}
  where the authors analyze cosmic inflation in such a bi-metric system.}
\end{center}
\index{bi-metric~spacetime}
\index{pseudo--Finsler~spacetime}
\end{table}

There is a subtlety implicit in setting the background velocities equal
that should be made explicit. If $\mathcal{V}_{1} =\mathcal{V}_{2} $
so that $\vec v_{\A0} = \vec v_{\B0}$, then since the masses appear in
the relationship between phase and velocity we deduce
\begin{equation}
m_\B \theta_{\A0}(t,\vec x) - m_\A \theta_{\B0}(t,\vec x) = f(t).
\end{equation}
If $m_\A\neq m_\B$, and if the background velocity is nonzero, we must
deduce that $\delta_{\A\B}(t,x)$ will be at the very least be position
dependent, and we will be unable to set it to zero. Alternatively, if
we demand $\delta_{\A\B}=0$, and have $ \nabla \theta_{\A0}(t,\vec x)=
\nabla \theta_{\B0}(t,\vec x) \neq 0$, then we cannot set $\vec
v_{\A0} = \vec v_{\B0} \neq 0$. Fortunately this will not seriously
affect further developments.

Last, but certainly not least, we present the conditions for a
mono-metric \index{mono-metric~spacetime} geometry in a 2-component BEC.

%+++++++++++++++++++++++++++++++++++++++++++++++++++++++++++++++++
\subsection{Mono-metric geometry \label{mono-metric}}
%+++++++++++++++++++++++++++++++++++++++++++++++++++++++++++++++++

Despite the fact that there are three different routes to
bi-metricity, once one demands mono-metricity, where
\begin{equation}
\mathbf{f}^{ab} =\diag \left[ f^{ab}_{11},f^{ab}_{11} \right]
= \diag \left[ \sqrt{-g_{11}}\, g^{ab}_{11},\sqrt{-g_ {11}}\, g^{ab}_{11} \right],
\end{equation}
then one ends up with one set unique of constraints to reduce from
pseudo--Finsler \index{pseudo--Finsler~spacetime} to a single-metric
Lorentzian geometry, namely:
\begin{equation}
\begin{split}
&\vec{v}_{\A0} = \vec{v}_{\B0} = \vec v_0; \\
&\tilde{U}_{\A\B}=0 ;\\
&\tilde{U}_{\A\A}=\tilde{U}_{\B\B} = \tilde U ; \\
&d_{\A} = d_{\B} = d.
\end{split}
\end{equation} 
This tuning completely specifies the spacetime geometry, in that
\begin{equation}
f_{11}^{ab}=f_{22}^{ab} \propto
\left(
\begin{array}{c|c}
- 1 & - \vec{v}_{0}^{\T} \\
\hline
-\vec{v}_{0} & \tilde{U}  d \, h^{ij} -  \vec{v}_{0}\vec{v}_{0}^{\T}  ,
\end{array}
\right)
\end{equation}
and after a small calculation we get
\begin{equation}
{g_{11}}^{ab}={g_{22}}^{ab}\propto
\left(
\begin{array}{c|c}
-(c^{2} - v_{0}^{2}) & - \vec{v}_{0}^{\T} \\
\hline
-\vec{v}_{0} &  h^{ij} 
\end{array},
\right)
\end{equation}
where we have defined
\begin{equation}
c^{2} = \tilde{U} \, d,
\end{equation}
as the speed of sound.\footnote{The speed of sound for quasi-particle
  excitations is of course our analogue for the speed of light in real
  gravity.}\index{quasi-particle}

Throughout the preceding few pages we have analyzed in detail the
first term in equation (\ref{E:fab_general}), and identified different
condensate parameters with different emergent
\index{emergent~geometry} geometries. Since there is more then one
term in the wave \index{wave~equation} equation describing excitations
in a two-component system, this is not the end of the story.  The
remaining terms in equation (\ref{E:fab_general}), which we might
generically view as ``mass'' and ``vector potential'' terms, do not
directly affect the spacetime geometry as such.  But when an
excitation propagates through a specific analogue
spacetime\index{analogue~spacetime} geometry, these terms will
contribute to the kinematics. It then becomes useful to consider the
``mass eigenmodes'' \index{eigenmodes} in field-space.

%+++++++++++++++++++++++++++++++++++++++++++++++++++++++++++++++++
\subsection{Merging spacetime geometry with mass eigenmodes \label{MergingGeoEig}}
%+++++++++++++++++++++++++++++++++++++++++++++++++++++++++++++++++

The eigenmodes \index{eigenmodes} we are interested in are eigenmodes
of the field-space matrices occurring in the sub-dominant terms of the
wave \index{wave~equation} equation. These eigenmodes (when they
exist) do not notice the presence of multiple fields --- in our
specific case a 2-field system --- and therefore propagate nicely
through the effective curved spacetime.  As promised in the abstract
and the motivation, we are striving for an analogue model representing
a massive scalar field in a mono-metric \index{mono-metric~spacetime}
Lorentzian structure.  By using the results from section
\ref{mono-metric} we are able to decouple the first term of equation
(\ref{E:fab_general}).

In the following we are focusing on two issues: First, we decouple the
remaining terms in equation (\ref{E:fab_general}), and subsequently we
check that these eigenmodes \index{eigenmodes} do not recouple the
geometric term.  There is however one more (technical) problem, and
that is the fact that the terms we want to associate with the
effective mass of the scalar field still contain partial derivatives
in time and space, which ultimately implies a dependence on the energy
of the propagating modes.\footnote{This can be easily be seen by going
  to the eikonal\index{eikonal} approximation where
  $\partial_{\vec{x}}\to i\vec{k}$ and $\partial_{t}\to i\omega$.}
Luckily, this problem can be easily circumvented, for equal background
phases,\footnote{Note that $\delta_{\A\B}=0$ plus mono=metricity
  implies either $m_\A=m_\B$ with arbitrary $\vec v_0\neq 0$, or
  $m_\A\neq m_\B$ with zero $\vec v_0= 0$. These are exactly the two
  situations we shall consider below.}
\begin{equation}
\theta_{\A0} = \theta_{\B0},
\end{equation}
in which case
\begin{equation}
\K=\G^{t}=\G^{i}=0.
\end{equation}
This has the effect of retaining only the matrix $\Lambda$ among the
sub-dominant terms, so that the wave equation \index{wave~equation} becomes
\begin{eqnarray} 
\partial_a ( \mathbf{f}^{ab} \partial_b \bar\theta) 
 + \Lambda  \; \bar{\theta} =0.
\end{eqnarray}
Due to the fact that the structure of the coupling matrix $\Lambda$
cannot be changed, its eigenmodes \index{eigenmodes} determine the
eigenmodes of the overall wave \index{wave~equation} equation. The
eigenvectors of $\Lambda$ are given by
\begin{equation}
\begin{split}
\mathrm{EV1} &:= [+1,+1] \\
\mathrm{EV2} &:= [-1,+1]
\end{split}
\end{equation}
The final step is to make sure that our spacetime geometry commutes
with the eigenvectors \index{eigenvectors} of $\Lambda$, that is
\begin{equation}
\left[ \mathbf{f}^{ab},\Lambda \right] = 0.
\end{equation}
This constraint is only fulfilled in the mono-metric
\index{mono-metric~spacetime} case, where we are dealing with two
identical classical fields, that effectively do not communicate with
each other.\footnote{While $\tilde{U}_{\A\B}=0$, $U_{\A\B}\neq 0$.}
That is, all field matrices are proportional to the identity matrix.

%+++++++++++++++++++++++++++++++++++++++++++++++++++++++++++++++++
\subsection{Special case:  $\Xi=constant$.}
%+++++++++++++++++++++++++++++++++++++++++++++++++++++++++++++++++

There is one specific class of geometries we are particularly
interested it, and that is when $\Xi$ is a position independent and
time independent constant. In the next section we will focus
exclusively on this case, and apply it to quantum
gravity\index{quantum~gravity!phenomenology} phenomenology.  This case
is however, also of interest as an example of an alternate interplay
between fine tuning and emergent geometry.\index{emergent~geometry}
Under the assumption that $\Xi$ is position and time independent, we
are able to directly manipulate the overall wave equation
\index{wave~equation} for the excitations and as a consequence obtain
slightly milder tuning conditions for mono-metricity.

Let us define
\begin{equation} \label{tildephase}
\tilde\theta = \Xi^{-1/2}\; \bar\theta,
\end{equation}
and multiply the whole wave equation (\ref{E:fab_general}) with
$\Xi^{1/2}$ from the left. What we are doing is a transformation in
field space onto a new basis $\tilde{\theta}$, and in the new basis the
wave \index{wave~equation} equation is given by,
\begin{equation}
\label{E:fab_general_constXi}
\partial_a \left( \tilde{\mathbf{f}}^{ab} \; \partial_b  \tilde\theta  \right)
+ \left(\tilde{\Lambda} +  \tilde{\K} \right)  \; \tilde{\theta}
+ {1\over2}\left\{
\tilde{\G}^{a} \; \partial_{a} {\tilde{\theta}}  + \partial_a (\tilde {\G}^a \tilde\theta)
\right\}= 0,
\end{equation}
where the matrices in field space transform as:
$\tilde\Lambda=\Xi^{1/2} \Lambda \Xi^{1/2}$, $\tilde\K=\Xi^{1/2} \K
\Xi^{1/2}$, $\tilde {\G}^{a}= \Xi^{1/2} \G^{a} \Xi^{1/2}$, and the
tensor-density as
\begin{equation}
\tilde{\mathbf{f}}^{ab} = \Xi^{1/2} \, \mathbf{f}^ {ab} \,\Xi^{1/2}.
\end{equation}
In general, the transformation matrix $\Xi^{1/2}$ is a non-diagonal,
though always symmetric:\footnote{See appendix B.}
\begin{equation}
\Xi^{1/2} =
\frac{\Xi +\sqrt{\det \Xi} \;\; \mathbf{I} }{\sqrt{\tr[\Xi]+ 2  \sqrt {\det \Xi}}}.
\end{equation}
A close look at equation (\ref{Finsler_product}), now using the
tensor-density $\tilde{\mathbf{f}}^{ab}$, makes it obvious that for
\begin{equation}
\tilde{U}_{\A\B} = 0,
\end{equation}
the geometry reduces from pseudo-Finsler to bi-metric.\index{bi-metric~spacetime} 
For the sake of keeping the discussion short and easy to
follow, we set the background velocities equal, and now get
\begin{equation}
\label{bi-metric_product_constXi}
\tilde{\mathbf{f}}^{a b} =
\mathcal{V}^{a} {\mathcal{V}^{b}} + \tilde{D}  h^{a b}.
\end{equation}
In view of the tuning, $\tilde{U}_{\A\B}=0$, we see
\begin{equation}
\tilde{D} = \diag (\tilde{U}_{\A\A}\, d_{\A},\tilde{U}_{\B\B}\, d_{\B}).
\end{equation}
The new mass-density matrix, and therefore the overall geometry is
diagonal in field space, hence we are now dealing with the required
bi-metric structure.

So far we are in complete agreement with what we have obtained in our
previous analysis, see Fig. \ref{Geometry}.  However, if we now ask
for mono-metricity, \index{mono-metric~spacetime} we obtain a slightly
milder constraint:
\begin{equation}
\tilde{U}_{\A\A}\, d_{\A} = \tilde{U}_{\B\B}\, d_{\B}.
\end{equation}

Last but not least, we show in detail the results we obtain for this
tuning scenario when including the $\Lambda$ term (the mass term). To
avoid confusion, we re-define a few matrices,
\begin{equation} \label{Omega2}
C_{0}^2 = \Xi^{1/2}\; D \;\Xi^{1/2} \, ;
\qquad \hbox{and} \qquad
\Omega^2 =   \Xi^{1/2} \;\Lambda\; \Xi^{1/2}.
\end{equation}
Both $C_{0}^2$ and $\Omega^2$ are symmetric matrices. If $ [C_{0}^2,
\; \Omega^2] = 0$, which is equivalent to the matrix equation $D \;
\Xi \; \Lambda = \Lambda \; \Xi \; D$, and is certainly satisfied in
view of the above constraint, then they have common eigenvectors. \index{eigenvectors}
Decomposition onto the eigenstates of the system results in a pair of
independent Klein--Gordon \index{Klein--Gordon equation} equations
\begin{equation} \label{KGE}
\frac{1}{\sqrt{-g_{\mathrm{I/II}}}}\,
\partial_{a} \left\{   \sqrt{-g_{\mathrm{I/II}}} \; (g_{\mathrm{I/ II}})^{ab} \;
\partial_{b} \tilde{\theta}_{\mathrm{I/II}} \right\} +
\omega_{\mathrm{I/II}}^2 \;
\tilde{\theta}_{\mathrm{I/II}} = 0 \; ,
\end{equation}
where the ``acoustic metrics'' are given by
\begin{equation} \label{metric}
(g_{\mathrm{I/II}})_{ab} \propto
\left[
\begin{array}{ccc}
-\left( c^2-v_0^2 \right)       &|& -\vec{v_0}^{\,T} \\
\hline
-\vec{v_0}  &|& \mathbf{I}_{d\times d}
\end{array}
\right].
\end{equation}
The metric components depend only on the background velocity
$\vec{v}_{0}$ and the common speed of sound $c$.\; It is also possible
to calculate the eigenfrequencies \index{eigenfrequencies} of the two
phonon\index{phonon} modes,
\begin{equation}
\omega_{I}^2 = 0;
\qquad
\omega_{II}^2 = \tr[\Omega^2] \, .
\end{equation}
A zero/ non-zero eigenfrequency \index{eigenfrequencies} corresponds
to a zero/ non-zero mass for the phonon\index{phonon} mode. %

In the eikonal\index{eikonal} limit we see that the in-phase perturbation will
propagate with the speed of sound,
\begin{equation}
\vec{v}_{s} =\vec{v}_0 + \hat k \;c\, ,
\end{equation}
while the anti-phase perturbations propagates with a lower group velocity given by:
\begin{equation}
\vec{v}_{g}= \frac{\partial \omega}{\partial \vec{k}}
=\vec{v}_0 + \hat k \;\frac{c^2}{\sqrt{\omega_{II}^2\; + c^2\; k^2}}.
\end{equation}
Here $k$ is the usual wave number. The dispersion relation we obtain
for the mono-metric \index{mono-metric~spacetime} structure is Lorentz
invariant.\index{Lorentz~invariance}

The fact that we have an analogue model representing both massive and
massless particles is promising for quantum gravity
phenomenology\index{quantum~gravity!phenomenology} if we now extend
the analysis to high-energy phonon\index{phonon} modes where the
quantum \index{quantum~pressure} pressure term is significant, and
where we consequently expect a breakdown of Lorentz
\index{Lorentz~invariance} invariance.  For the following, we
concentrate on the generalization of flat Minkowski spacetime, which
implies a constant $\Xi$ and zero background velocities, $\vec{v}_0$.
In the language of condensed matter physics, we are thinking of a
uniform condensate at rest.

%%%%%%%%%%%%%%%%%%%%%%%%%%%%%%%%%%%%%%%%%%%%%%%%%%%

\section{Application to  quantum gravity phenomenology}                                   

%%%%%%%%%%%%%%%%%%%%%%%%%%%%%%%%%%%%%%%%%%%%%%%%%%%

In using this 2-BEC \index{2-BEC~system} model to probe issues of
interest to the ``quantum gravity\index{quantum~gravity!phenomenology}
phenomenology'' community it behooves us to simplify as much as
possible the parts of the model not of direct interest for current
considerations. Specifically, we wish to use the ``quantum
\index{quantum~pressure} pressure'' term as a model for the type of
Lorentz \index{Lorentz~invariance!violation} violating physics that
might occur in the physical universe at or near the Planck
\index{Planck~scale} scale~\cite{Breakdown}. Since we are then interested in high
energies, and consequently short distances, one might expect the
average spacetime curvature to be negligible --- that is, we will be
interested in looking for ``quantum \index{quantum~pressure}
pressure'' induced deviations from special
relativity,\index{special~relativity} and can dispense with the notion
of curved spacetimes for now. (``Flat'' pseudo--Finsler
\index{pseudo--Finsler~spacetime} spaces are already sufficiently
complicated to lead to interesting physics.)  In terms of the BEC
condensates this means that in this section of the chapter we will
concentrate on a spatially-homogeneous time-independent background, so
that in particular all the matrices $\mathbf{f}^{ab}$ will be taken to
be position-independent. (And similarly, $\Xi$, $\Lambda$, $D$,
\emph{etc.} are taken to be position independent and we set $\vec
v_0=0$, so the background is at rest.)  This greatly simplifies the
calculations (though they are still relatively messy), but without
sacrificing the essential pieces of the physics we are now interested
in.

%
%_figure__figure__figure__figure__figure__figure__figure__figure__figure__figure__figure__figure_
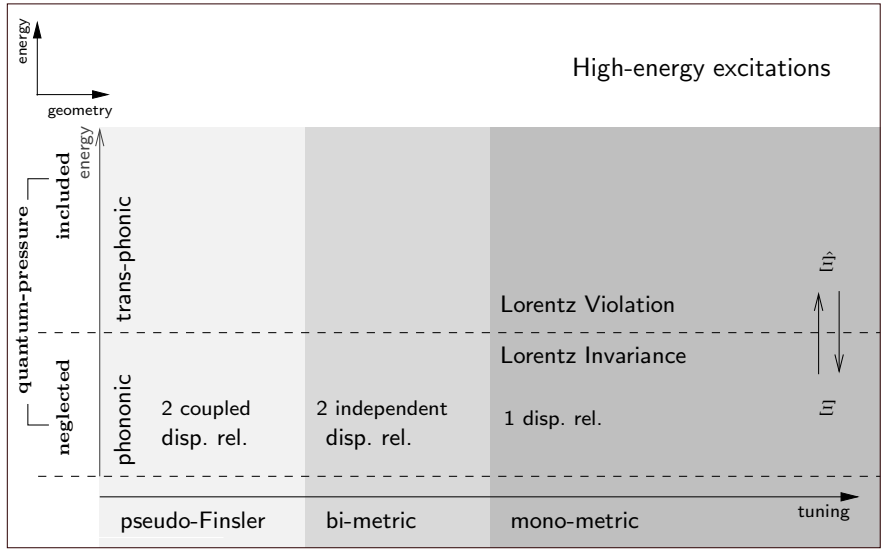
\begin{figure}[htb]
 \begin{center}
 \input{UV_Physics.pstex_t}
 \caption[Concept for fine tuning.]  {\label{Concept}How to tune the
   system to exhibit breakdown of Lorentz symmetry.}
 \end{center} 
\index{Lorentz~invariance}      
\end{figure}
%_figure__figure__figure__figure__figure__figure__figure__figure__figure__figure__figure__figure_
%

Now the purpose of quantum
gravity\index{quantum~gravity!phenomenology} phenomenology is to
analyze the physical consequences arising from various models of
quantum gravity.  One hope for obtaining an experimental grasp on
quantum gravity is the generic prediction arising in many (but not
all) quantum gravity\index{quantum~gravity} models that ultraviolet
physics at or near the Planck scale,\index{Planck~scale}
$M_{\mathrm{Planck}} = 1.2 \times 10^{19} \; \mathrm{GeV/c^2}$, (or in
some models the string scale), typically induces violations of Lorentz
\index{Lorentz~invariance!violation} invariance at lower
scales~\cite{LIV, jlm-ann}.  Interestingly most investigations, even
if they arise from quite different fundamental physics, seem to
converge on the prediction that the breakdown of Lorentz
\index{Lorentz~invariance} invariance can generically become manifest
in the form of modified dispersion relations
\begin{equation}
 \label{disp}
\omega^2 = \omega_0^2 + \left(1 + \eta_{2} \right) \, c^2 \; k^2 
+ \eta_{4} \, \left(\frac{\hbar}{M_{\mathrm{Lorentz~violation}}} \right)^2  \; k^4 + \dots \; ,
\end{equation}
where the coefficients $\eta_{n}$ are dimensionless (and possibly
dependent on the particle species considered), and we have restricted
our expansion to CPT\index{CPT~invariance} invariant terms (otherwise
one would also get odd powers in $k$). The particular inertial frame
for these dispersion relations is generally specified to be the frame
set by cosmological microwave background, and $M_{\mathrm{Lorentz~violation}}$
is the scale of Lorentz \index{Lorentz~invariance} symmetry breaking
which furthermore is generally assumed to be of the order of
$M_{\mathrm{Planck}}$.\index{Planck~scale}

Although several alternative scenarios have been considered in the
literature in order to justify the modified kinematics discussed
above, to date the most commonly explored avenue is an effective field
theory (EFT)\index{effective~field~theory} approach.  In the present
chapter we focus on the class of non-renormalizable
\EFTs\index{effective~field~theory} with Lorentz
\index{Lorentz~invariance!violation} violations associated to
dispersion relations like equation~(\ref{disp}).  Relaxing our
CPT\index{CPT~invariance} invariance condition this class would
include the model developed in~\cite{MP}, and subsequently studied by
several authors, where an extension of quantum electrodynamics including only mass
dimension five Lorentz-violating operators was considered. (That
ansatz leads to order $k^3$ Lorentz
\index{Lorentz~invariance!violation} and CPT\index{CPT~violation}
violating terms in the dispersion relation.)  Very accurate
constraints have been obtained for this model using a combination of
experiments and observations (mainly in high energy astrophysics). See
\emph{e.g.}\/~\cite{jlm-ann,Jacobson:2002hd,Jacobson:Limit,Jacobson:Crab}.
In spite of the remarkable success of this framework as a ``test
theory'', it is interesting to note that there are still significant
open issues concerning its theoretical foundations.  Perhaps the most
pressing one is the so called {\em naturalness problem} which can be
expressed in the following way: Looking back at our ansatz
(\ref{disp}) we can see that the lowest-order correction, proportional
to $\eta_{2}$, is not explicitly Planck suppressed.\index{Planck
  suppressed} This implies that such a term would always be dominant
with respect to the higher-order ones and grossly incompatible with
observations (given that we have very good constraints on the
universality of the speed of light for different elementary
particles).  Following the observational leads it has been therefore
often assumed either that some symmetry (other than Lorentz
\index{Lorentz~invariance} invariance) enforces the $\eta_2$
coefficients to be exactly zero, or that the presence of some other
characteristic \EFT\index{effective~field~theory} mass scale $\mu\ll
M_{\Pl}$ (\emph{e.g.}, some particle physics mass scale)
associated with the Lorentz \index{Lorentz~invariance} symmetry
breaking might enter in the lowest order dimensionless coefficient
$\eta_{2}$ --- which will be then generically suppressed by
appropriate ratios of this characteristic mass to the Planck
\index{Planck~scale} mass: $\eta_2\propto (\mu/M_{\Pl})^\sigma$
where $\sigma\geq 1$ is some positive power (often taken as one or
two). If this is the case then one has two distinct regimes: For low
momenta $p/(M_{\Pl}c) \ll (\mu/M_{\Pl})^\sigma$ the lower-order
(quadratic in the momentum) deviations in~(\ref{disp}) will dominate
over the higher-order ones, while at high energies $p/(M_{\Pl}c)
\gg (\mu/M_{\Pl})^\sigma$ the higher order terms will be dominant.

The naturalness problem arises because such a scenario is not well
justified within an \EFT\index{effective~field~theory} framework; in
other words there is no natural suppression of the low-order
modifications in these models. In fact we implicitly assumed that
there are no extra Planck\index{Planck suppressed} suppressions hidden
in the dimensionless coefficients $\eta_n$ with $n>2$.
\EFT\index{effective~field~theory} cannot justify why \emph{only} the
dimensionless coefficients of the $n\leq 2$ terms should be suppressed
by powers of the small ratio $\mu/M_{\Pl}$.  Even worse,
renormalization group arguments seem to imply that a similar mass
ratio, $\mu/M_{\Pl}$ would implicitly be present also in \emph{all}
the dimensionless $n>2$ coefficients --- hence suppressing them even
further, to the point of complete undetectability.  Furthermore it is
easy to show~\cite{Collins} that, without some protecting symmetry, it
is generic that radiative corrections due to particle interactions in
an \EFT\index{effective~field~theory} with only Lorentz
\index{Lorentz~invariance!violation} violations of order $n>2$ in
(\ref{disp}) for the free particles, will generate $n= 2$ Lorentz
\index{Lorentz~invariance!violation} violating terms in the dispersion
relation, which will then be dominant.  Observational
evidence~\cite{LIV} suggests that for a variety of standard model
particles $|\eta_2|\lesssim 10^{-21}$. Naturalness in
\EFT\index{effective~field~theory} would then imply that the higher
order terms are at least as suppressed as this, and hence beyond
observational reach.

A second issue is that of ``universality'', which is not so much a
``problem'', as an issue of debate as to the best strategy to adopt.
In dealing with situations with multiple particles one has to choose
between the case of universal (particle-independent) Lorentz
\index{Lorentz~invariance!violation} violating coefficients $\eta_n$,
or instead go for a more general ansatz and allow for
particle-dependent coefficients; hence allowing different magnitudes
of Lorentz \index{Lorentz~invariance!violation} symmetry violation for
different particles even when considering the same order terms (same
$n$) in the momentum expansion. The two choices are equally
represented in the extant literature (see \emph{e.g.}~\cite{GAC-Pir}
and \cite{Jacobson:2002hd} for the two alternative ans\"atze), but it
would be interesting to understand how generic this universality might
be, and what sort of processes might induce non-universal Lorentz
\index{Lorentz~invariance!violation} violation for different
particles.

%+++++++++++++++++++++++++++++++++++++++++++++++++++++++++++++++++
\subsection{Specializing the wave equation\label{sec:wave-equation}}%
%+++++++++++++++++++++++++++++++++++++++++++++++++++++++++++++++++

For current purposes, where we wish to probe violations of Lorentz
\index{Lorentz~invariance!violation} invariance in a flat analogue
spacetime,\index{analogue~spacetime} we start with our basic wave
equation (\ref{Eq:GHWE}) and make the following specializations:
$\delta_{\A\B}\to0$ (so that $\G^a\to0$ and $K\to0$). We also set all
background fields to be homogeneous (space and time independent), and
use the formal operators $\hat \Xi^{1/2}$ and $\hat \Xi^{-1/2}$ to
define a new set of variables
\begin{equation}
\tilde\theta = \hat \Xi^{-1/2} \;\bar\theta,
\end{equation}
in terms of which the wave \index{wave~equation} equation becomes
\begin{equation}
\label{E:simplified-wave-equation}
\partial_t^2\tilde\theta = 
\left\{ \hat \Xi^{1/2}\; [ D \nabla^2 - \Lambda] \; \hat \Xi^{1/2} \right\} \tilde\theta,
\end{equation}
or more explicitly
\begin{equation}
\label{eq:last}
\partial_t^2\tilde\theta = 
\left\{ [\Xi- X \nabla^2]^{1/2}\;  [ D \nabla^2 - \Lambda] \; 
 [\Xi-X\nabla^2] ^{1/2} \right\} \tilde\theta.
\end{equation}
This is now a (relatively) simple PDE to analyze.  The objects $\hat
\Xi^{1/2}$ and $\hat \Xi^{-1/2}$ are $2\times2$ matrices whose
elements are pseudo-differential operators, but to simplify things it
is computationally efficient to go directly to the eikonal\index{eikonal} limit
where\footnote{Once we are in the eikonal\index{eikonal} approximation the
  pseudo-differential operator $\hat \Xi^{1/2} \to \sqrt{ \Xi + k^2 X}$
  can be given a simple and explicit meaning in terms of the
  Hamilton--Cayley theorems of appendix B.}
\begin{equation}
\hat \Xi \to \Xi + X\; k^2.
\end{equation}
This finally leads to a dispersion relation of the form
\begin{eqnarray}
\label{fresnel}
&&\det\big\{ \omega^2 \; \mathbf{I} -
 [\Xi+ X k^2]^{1/2}\;  [ D k^2  + \Lambda] \; 
 [\Xi+ X k^2] ^{1/2}  \big\} =0\,,
\end{eqnarray}
and ``all'' we need to do for the purposes of this chapter, is to
understand this quasiparticle excitation spectrum in detail.

%+++++++++++++++++++++++++++++++++++++++++++++++++++++++++++++++++
\subsection{Hydrodynamic approximation\label{sec:hydro}}
%+++++++++++++++++++++++++++++++++++++++++++++++++++++++++++++++++

The hydrodynamic \index{hydrodynamic~limit} limit consists of formally
setting $\hat X\to0$ so that $\hat \Xi \to \Xi$.  (That is, one is
formally setting the healing\index{healing~length} length matrix to
zero: $Y \to 0$. More precisely, all components of the healing length
matrix are assumed small compared to other length scales in the
problem.)  The wave \index{wave~equation} equation (\ref{eq:last}) now
takes the form:
\begin{eqnarray}
\partial_t^2\tilde\theta = 
\left\{ \Xi^{1/2}\; [ D \nabla^2 - \Lambda] \; \Xi^{1/2} \right\} \tilde\theta.
\end{eqnarray}
Since this is second-order in both space and time derivatives, we now
have at least the possibility of obtaining an exact ``Lorentz
\index{Lorentz~invariance} invariance''.  We can now define the
matrices
\begin{equation}
\Omega^2 = \Xi^{1/2}\;\Lambda\;\Xi^{1/2}; \qquad C_0^2 =   \Xi^{1/2}\; D\; \Xi^{1/2}; 
\end{equation}
so that after Fourier transformation
\begin{equation}
\omega^2 {\tilde{\theta}}  = \left\{ C_0^2 \;
k^2+\Omega^2\right\} \; \tilde{\theta}  \equiv H(k^2) \; 
\tilde{\theta},
\end{equation}
leading to the Fresnel\index{Fresnel~equation} equation
\begin{equation}
\det\{ \omega^2 \;\mathbf{I} - H(k^2) \} =0.
\end{equation}
That is
\begin{equation}
\omega^4 - \omega^2 \; \hbox{tr}[H(k^2)] + \det[H(k^2)] =
0,
\end{equation}
whence
\begin{equation}
\omega^2 = { \hbox{tr}[H(k^2)] \pm \sqrt{
    \hbox{tr}[H(k^2)]^2 - 4\; \det[H(k^2)] }\over 2}.
\label{eq:disp-rel-hydro}
\end{equation}
Note that the matrices $\Omega^2$, $C_0^2$, and $H(k^2)$ have now
carefully been arranged to be \emph{symmetric}. This greatly
simplifies the subsequent matrix algebra. Also note that the matrix
$H(k^2)$ is a function of $k^2$; this will forbid the appearance of
odd powers of $k$ in the dispersion relation --- as should be expected
due to the parity invariance of the system.

%~~~~~~~~~~~~~~~~~~~~~~~~~~~~~~~~~~~~~~~~~~~~~~~~~~~~
\subsubsection{Masses}
%~~~~~~~~~~~~~~~~~~~~~~~~~~~~~~~~~~~~~~~~~~~~~~~~~~~~
We read off the ``masses'' by looking at the special case of
space-independent oscillations for which
\begin{equation}
 \partial_{t}^2 {\bar{\theta}}  = -\Omega^2 \; \bar{\theta},
\end{equation}
allowing us to identify the ``mass'' (more precisely, the natural oscillation
frequency) as
\begin{equation}
\hbox{``masses''} 
\propto \hbox{eigenvalues of}\;(\Xi^{1/2}\;\Lambda\; \Xi^{1/2})
 =  \hbox{eigenvalues of}\;(\Xi\;\Lambda).
\end{equation}
Since $\Lambda$ is a singular $2\times2$ matrix this
automatically implies
\begin{equation}
\omega_I^2=0; \qquad \omega_{II}^2 = \hbox{tr}\,(\Xi\;\Lambda).
\label{eq:masses}
\end{equation}
So we see that one mode will be a massless phonon\index{phonon} while
the other will have a non zero mass. Explicitly, in terms of the
elements of the underlying matrices
\begin{equation}
\omega_I^2=0; \qquad  \omega_{II}^2 =
-\frac{2\sqrt{\rho_{\A0}\,\rho_{\B0}} \;\lambda }{\hbar^2}
\{ \tilde U_{\A\A} + \tilde U_{\B\B} - 2 \tilde U_{\A\B} \}
\label{eq:masses2}
\end{equation}
so that (before any fine-tuning or decoupling)
\begin{eqnarray}
\omega_{II}^2 &=& 
-\frac{2\sqrt{\rho_{\A0}\,\rho_{\B0}} \;\lambda}{\hbar^2} 
\\
&& \times \left\{ U_{\A\A} +  U_{\B\B} - 2 U_{\A\B}
-{\lambda\over2\sqrt{\rho_{\A0}\,\rho_{\B0}}}  \left[
  \sqrt{\rho_{\A0}\over \rho_{\B0}} 
+\sqrt{\rho_{\B0}\over
    \rho_{\A0}}\right]^2 \right\}.
\nonumber
\label{eq:m-ndiag}
\end{eqnarray}
It is easy to check that this quantity really does have the physical
dimensions of a frequency.

%~~~~~~~~~~~~~~~~~~~~~~~~~~~~~~~~~~~~~~~~~~~~~~~~~~~~
\subsubsection{Mono-metricity conditions\label{sec:mono-metr}}
%~~~~~~~~~~~~~~~~~~~~~~~~~~~~~~~~~~~~~~~~~~~~~~~~~~~~
In order for our system to be a perfect analogue of
special relativity:\index{special~relativity}
\begin{itemize}
\item we want each mode to have a quadratic dispersion
  relation;
\item we want each dispersion relation to have the same
  asymptotic slope.
\end{itemize}
Let us start by noticing that the dispersion relation
(\ref{eq:disp-rel-hydro}) is of the form
\begin{equation}
\omega^2 = [\hbox{quadratic}_1] \pm
\sqrt{[\hbox{quartic}]}.
\end{equation}
The first condition implies that the quartic must be a
perfect square
\begin{equation}
[\hbox{quartic}] = [\hbox{quadratic}_2]^2,
\end{equation}
but then the second condition implies that the slope of
this quadratic must be zero. That is
\begin{equation}
[\hbox{quadratic}_2](k^2) = [\hbox{quadratic}_2](0),
\end{equation}
and so
\begin{equation}
[\hbox{quartic}](k^2) =[ \hbox{quartic}](0) 
\end{equation}
must be constant independent of $k^2$, so that the
dispersion relation is of the form
\begin{equation}
\omega^2 = [\hbox{quadratic}_1](k^2) \pm[\hbox{quadratic}_2](0).
\end{equation}
Note that this has the required form (two hyperbolae with the same
asymptotes, and possibly different intercepts). Now let us implement
this directly in terms of the matrices $C_0^2$ and $M^2$.

\noindent\emph{Step 1:}
Using the results of the appendix, specifically equation (\ref{E:A-2-matrices}):
\begin{eqnarray}
\det[H^2(k)] &=& \det[\Omega^2 + C_0^2\; k^2 ] \\ &=&
\det[\Omega^2] - \tr\left\{ \Omega^2\; \bar C_0^2\right\} \; k^2 + \det[C_0^2]\;(k^2)^2.
\end{eqnarray}
(This holds for any linear combination of $2\times2$ matrices. Note
that we apply trace reversal to the squared matrix $C_0^2$, we do not
trace reverse and then square.)
Since in particular $\det[\Omega^2]=0$,  we have:
\begin{equation}
\det[H^2(k)] = - \tr\left\{ \Omega^2\;\bar C_0^2 \right\} \; k^2 + \det[C_0^2]\;(k^2)^2.
\end{equation}

\noindent\emph{Step 2:}
Now consider the discriminant (the quartic)
\begin{eqnarray}
 \qquad \hbox{quartic} &\equiv&
\hbox{tr}[H(k^2)]^2 - 4\; \det[H(k^2)] 
\\ 
 &=& (\tr[\Omega^2]+\tr[C_0^2]\; k^2)^2 - 4 \big[
    -\tr\left\{ \Omega^2\;\bar C_0^2 \right\} \; k^2 
\nonumber
\\
&&\qquad\qquad\qquad\qquad + \det[C_0^2]\;(k^2)^2 \big]
\qquad
  \\ 
 &=& \tr[\Omega^2]^2 + \{2 \tr[\Omega^2]\tr[C_0^2]+4
  \tr\left\{ \Omega^2\;\bar C_0^2 \right\} \} k^2 
\nonumber \\ &&\qquad\qquad\qquad\qquad
  + \left\{ \tr[C_0^2]^2 - 4\det[C_0^2]\right\} (k^2)^2
  \\
 &=& \tr[\Omega^2]^2 +
 2 \{2 \tr\left\{ \Omega^2\; C_0^2 \right\} - \tr[\Omega^2]\tr[C_0^2] \} k^2 
\nonumber \\ &&\qquad\qquad\qquad\qquad
  + \left\{ \tr[C_0^2]^2 - 4\det[C_0^2]\right\} (k^2)^2.
\end{eqnarray}
So in the end the two conditions above for mono-metricity
take the form
\begin{equation}
\hbox{mono-metricity}\iff\left\{ 
\begin{array}{l}
\tr[C_0^2]^2 - 4\;\det[C_0^2] =0;
\\
2 \tr\left\{ \Omega^2\; C_0^2 \right\} -  \tr[\Omega^2] \;\tr[C_0^2]= 0.
\end{array}\right .
\end{equation}
Once these two conditions are satisfied the dispersion
relation is
\begin{equation}
\omega^2 = { \hbox{tr}[H(k^2)] \pm \tr[\Omega^2] \over 2} =
      {\tr[\Omega^2]\pm\tr[\Omega^2] + \tr[C_0^2]\; k^2\over 2}
\end{equation}
whence
\begin{equation}
\omega_1^2 = {1\over2} \tr[C_0^2]\; k^2=c_0^2k^2
\qquad
\omega_2^2 = \tr[\Omega^2] + {1\over2} \tr[C_0^2]\; k^2=\omega_{II}^2+c_0^2k^2,
\end{equation}
as required. One mode is massless, one massive with
exactly the ``mass'' previously deduced. One can now define the quantity
\begin{equation}
m_{II} = \hbar \omega_{II}/c_0^2,
\end{equation}
which really does have the physical dimensions of a mass.

%~~~~~~~~~~~~~~~~~~~~~~~~~~~~~~~~~~~~~~~~~~~~~~~~~~~~
\subsubsection{Interpretation of the mono-metricity conditions\label{S:C1C2}}
%~~~~~~~~~~~~~~~~~~~~~~~~~~~~~~~~~~~~~~~~~~~~~~~~~~~~
But now we have to analyse the two simplification
conditions
\begin{eqnarray}
C1:&&\qquad \tr[C_0^2]^2 - 4\;\det[C_0^2]= 0;
\label{eq:monometr1}
\\
C2:&&\qquad 2\; \tr\left\{ \Omega^2\; C_0^2 \right\} - \tr[\Omega^2]\tr[C_0^2]= 0;
 \label{eq:monometr2}
\end{eqnarray}
to see what they tell us.  The first of these conditions is equivalent
to the statement that the $2\times2$ matrix $C_0^2$ has two identical
eigenvalues.\index{eigenvalues} But since $C_0^2$ is symmetric this
then implies $C_0^2 = c_0^2 \; \mathbf{I}$, in which case the second
condition is automatically satisfied. (In contrast, condition $C2$
does not automatically imply condition $C1$.)  Indeed if $C_0^2 =
c_0^2 \; \mathbf{I}$, then it is easy to see that (in order to make
$C_0^2$ diagonal)
\begin{equation}
\tilde U_{\A\B}=0,
\end{equation}
(which is sufficient, by itself, to imply bi-metricity) and
furthermore that
\begin{equation}
{\tilde U_{\A\A} \;\rho_{\A0}\over m_\A} = c_0^2 = {\tilde
  U_{\B\B} \;\rho_{\B0}\over m_\B}.
  \label{eq:c0ft}
\end{equation}
Note that we can now solve
for $\lambda$ to get
\begin{equation}
\lambda = -2 \sqrt{\rho_{\A0}\;\rho_{\B0}} \; U_{\A\B},
\label{eq:lbd}
\end{equation}
whence
\begin{equation}
c_0^2 = {U_{\A\A\;}\rho_{\A0}+U_{\A\B}\; \rho_{\B0}\over m_\A} =
{U_{\B\B\;}\rho_{\B0}+U_{\A\B}\; \rho_{\A0}\over m_\B},
\end{equation} 
and
\begin{equation}
\omega_{II}^2 = {4\rho_{\A0}\rho_{\B0}U_{\A\B}\over \hbar^2}  \left\{
U_{\A\A} + U_{\B\B} - 2U_{\A\B} + U_{\A\B}  \left[
  \sqrt{\rho_{\A0}\over \rho_{\B0}} +\sqrt{\rho_{\B0}\over
    \rho_{\A0}}\right]^2 \right\}.
\label{eq:m-diag}
\end{equation}
Note that (\ref{eq:m-diag}) is equivalent to
(\ref{eq:m-ndiag}) with (\ref{eq:lbd}) enforced. But this then implies
\begin{equation}
\label{E:m2}
\omega_{II}^2 = {4\rho_{\A0}\rho_{\B0}U_{\A\B}\over \hbar^2}  \left\{
U_{\A\A} + U_{\B\B} + U_{\A\B}  \left[
  {\rho_{\A0}\over \rho_{\B0}} +{\rho_{\B0}\over
    \rho_{\A0}}\right] \right\}.
\end{equation}
\emph{Interpretation:} Condition $C2$ forces the two low-momentum
``propagation speeds'' to be the same, that is, it forces the two
$O(k^2)$ coefficients to be equal. Condition $C1$ is the stronger
statement that there is no $O(k^4)$ (or higher order) distortion to
the relativistic dispersion relation.

%+++++++++++++++++++++++++++++++++++++++++++++++++++++++++++++++++
\subsection{Beyond the hydrodynamical approximation \label{sec:qp}}
%+++++++++++++++++++++++++++++++++++++++++++++++++++++++++++++++++

At this point we want to consider the deviations from the previous
analogue for special relativity.\index{special~relativity} Our
starting point is again equation (\ref{eq:last}), now retaining the
quantum \index{quantum~pressure} pressure term, which we Fourier
transform to get:
\begin{equation}
\omega^2 {\tilde{\theta}}  = 
\left\{
\sqrt{\Xi+X\; k^2} \;\; [D\; k^2+\Lambda]\;\; \sqrt{\Xi+X\;k^2} 
\right\}\;  \tilde{\theta}  \equiv H(k^2) \; \tilde{\theta}.
   \label{eq:new-disp-rel}
\end{equation}
This leads to the Fresnel\index{Fresnel~equation} equation
\begin{equation}
\det\{ \omega^2 \;\mathbf{I} - H(k^2) \} =0.
\end{equation}
That is
\begin{equation}
\omega^4 - \omega^2 \; \hbox{tr}[H(k^2)] + \det[H(k^2)] =
0,
\end{equation}
whence
\begin{equation}
\omega^2 = { \hbox{tr}[H(k^2)] \pm \sqrt{
    \hbox{tr}[H(k^2)]^2 - 4\;  \det[H(k^2)] }\over 2},
\label{eq:tot-disp-rel}
\end{equation}
which is now of the form
\begin{equation}
\omega^2 = [\hbox{quartic}_1] \pm \sqrt{[ \hbox{octic} ]}.
\end{equation}

%~~~~~~~~~~~~~~~~~~~~~~~~~~~~~~~~~~~~~~~~~~~~~~~~~~~~
\subsubsection{Masses}
%~~~~~~~~~~~~~~~~~~~~~~~~~~~~~~~~~~~~~~~~~~~~~~~~~~~~
The ``masses'', defined as the zero momentum oscillation frequencies,
are again easy to identify. Just note that the $k$-independent term in
the Fresnel\index{Fresnel~equation} equation is exactly the same mass matrix
$\Omega^2=\Xi^{1/2}\;\Lambda\;\Xi^{1/2}$ that was present in the
hydrodynamical limit. (That is, the quantum potential term $X$
does not influence the masses.)

%~~~~~~~~~~~~~~~~~~~~~~~~~~~~~~~~~~~~~~~~~~~~~~~~~~~~
\subsubsection{Dispersion relations}
%~~~~~~~~~~~~~~~~~~~~~~~~~~~~~~~~~~~~~~~~~~~~~~~~~~~~
Differently from the previous case, when the hydrodynamic
\index{hydrodynamic~limit} approximation held, we now have that the
discriminant of (\ref{eq:tot-disp-rel}) generically can be an
eighth-order polynomial in $k$.  In this case we cannot hope to
recover an exact analogue of special
relativity,\index{special~relativity} but instead can at best hope to
obtain dispersion relations with vanishing or {\em suppressed}
deviations from special relativity at low $k$; possibly with large
deviations from special relativity at high momenta. From the form of
our equation it is clear that the Lorentz
\index{Lorentz~invariance!violation} violation suppression should be
somehow associated with the masses of the atoms $m_{A/B}$. Indeed we
will use the underlying atomic masses to define our ``Lorentz
\index{Lorentz~invariance!violation} breaking scale'', which we shall
then assume can be identified with the ``quantum gravity scale''.  The
exact form and relative strengths of the higher-order terms will be
controlled by tuning the 2--BEC system and will eventually decide the
manifestation (or not) of the naturalness problem and of the
universality issue.

Our approach will again consist of considering derivatives of
(\ref{eq:tot-disp-rel}) in growing even powers of $k^2$ (recall that
odd powers of $k$ are excluded by the parity invariance of the system)
and then setting $k\to0$. We shall compute only the coefficients up to
order $k^4$ as by simple dimensional arguments one can expect any
higher order term will be further suppressed with respect to the $k^4$
one.

We can greatly simplify our calculations if before performing our
analysis we rearrange our problem in the following way. First of all
note that by the cyclic properties of trace
\begin{eqnarray}
\tr[H(k^2)] &=& \tr[ (Dk^2+\Lambda)\;(\Xi+k^2X)] 
\\
&=& \tr[ \Lambda \Xi + k^2(D\Xi+\Lambda X) + (k^2)^2 D X]
\\
&=& \tr[  \Xi^{1/2} \Lambda \Xi^{1/2} + k^2(\Xi^{1/2} D \Xi^{1/2} 
+ X^{1/2}\Lambda X^{1/2})
\nonumber
\\
&&
+ (k^2)^2 X^{1/2} D X^{1/2}].
\end{eqnarray}
Putting this all together, we can now define symmetric matrices
\begin{equation}
\Omega^2 =   \Xi^{1/2} \Lambda \Xi^{1/2}; 
\end{equation}
\begin{equation}
C_0^2 = \Xi^{1/2} D \Xi^{1/2}; \qquad
\Delta C^2 =  X^{1/2}\Lambda X^{1/2};
\end{equation}
\begin{equation}
 C^2 = C_0^2 + \Delta C^2 = \Xi^{1/2} D \Xi^{1/2} + X^{1/2}\Lambda X^{1/2}; 
\end{equation}
\begin{equation}
Z^2 =  2 X^{1/2} D X^{1/2} = {\hbar^2\over 2} M^{-2}.
\end{equation}
With all these definitions we can then write
\begin{equation}
\tr[H(k^2)] = \tr\left[ \Omega^2 + k^2 (C_0^2+\Delta C^2) + {1\over2} (k^2)^2 Z^2 \right],
\end{equation}
where everything has been done inside the trace. If we now define
\begin{equation}
H_s(k^2) = \Omega^2 + k^2 (C_0^2+\Delta C^2) + {1\over2} (k^2)^2 Z^2,
\end{equation}
then $H_s(k^2)$ is by definition both polynomial and symmetric and satisfies
\begin{equation}
\tr[H(k^2)] = \tr[H_s(k^2)],
\end{equation}
while in contrast, 
\begin{equation}
\det[H(k^2)] \neq \det[H_s(k^2)].
\end{equation}
But then
\begin{equation}
\omega^2 = {1\over2}\left[ \tr[H_s(k^2)] \pm \sqrt{ \tr[H_s(k^2)]^2-4\det[H(k^2)]}\right].
\end{equation}
Whence
\def\d{{\mathrm{d}}}
\begin{equation}
{\d\omega^2\over\d k^2} = {1\over2}\left[\tr[H_s'(k^2)] \pm 
{
\tr[H_s(k^2)]\tr[H_s'(k^2)] - 2\det'[H(k^2)] \over
\sqrt{ \tr[H_s(k^2)]^2- 4\det[H(k^2)] }
}\right],
\end{equation}
and at $k=0$
\begin{equation}
\left.{\d\omega^2\over\d k^2}\right|_{k\to0} = {1\over2}\left[\tr[C^2] \pm 
{
\tr[\Omega^2]\tr[C^2] - 2\det'[H(k^2)]_{k\to0} \over
 \tr[\Omega^2] 
}\right].
\end{equation}
But now let us consider 
\begin{eqnarray}
\det[H(k^2)] &=& \det[  (Dk^2+\Lambda)\;(\Xi+k^2X) ] 
\\
&=& 
\det[Dk^2+\Lambda]\;\det[\Xi+k^2X] 
\\
&=&
\det[ \Xi^{1/2} (Dk^2+\Lambda) \Xi^{1/2}] \;
\det[ I + k^2 \Xi^{-1/2} X \Xi^{-1/2} ] 
\end{eqnarray}
where we have repeatedly used properties of the determinant.
Furthermore
\begin{eqnarray}
 \det[ I + k^2 \Xi^{-1/2} X \Xi^{-1/2} ] &=& \det[I+k^2 \Xi^{-1} X] 
 \\
&=& \det[I+k^2 X^{1/2} \Xi X^{1/2}] 
 \\
 &=& \det[I+k^2 Y^2/2],
\end{eqnarray}
so that we have
\begin{equation}
\det[H(k^2)] = \det[\Omega^2+C_0^2 k^2] \; \det[I + k^2 Y^2/2].
\end{equation}
Note the the matrix $Y^2$ is the ``healing\index{healing~length}
length matrix'' we had previously defined, and that the net result of
this analysis is that the full determinant is the product of the
determinant previously found in the hydrodynamic
\index{hydrodynamic~limit} limit with a factor that depends on the
product of wavenumber and healing\index{healing~length} length.

But now, given our formula (\ref{E:A-2-matrices}) for the determinant,
we see
\begin{eqnarray}
{\det}'[H(k^2)] &=& (-\tr(\Omega^2\bar C_0^2)+ 2 k^2 \det[C_0^2] ) \; \det[I + k^2 Y^2/2] 
\nonumber
\\
&&
+  \det[\Omega^2+C_0^2 k^2] \; (-\tr[\bar Y^2] +  k^2 \det[Y^2])/2,
\end{eqnarray}
whence
\begin{equation}
{\det}'[H(k^2)]_{k\to0} = -\tr(\Omega^2\bar C_0^2 ),
\end{equation}
and so
\begin{equation}
\left.{\d\omega^2\over\d k^2}\right|_{k\to0} = {1\over2}\left[\tr[C^2] \pm 
{
\tr[\Omega^2]\tr[C^2] + 2 \tr(\Omega^2\bar C_0^2 ) \over
 \tr[\Omega^2] 
}\right].
\end{equation}
That is:
\begin{equation}
\label{eq: varpi2}
\left.{\d\omega^2\over\d k^2}\right|_{k\to0} = 
{1\over2}\left[\tr[C^2] 
\pm 
\left\{ \tr[C^2] + 2{ \tr(\Omega^2\bar C_0^2 ) \over \tr[\Omega^2] }
\right\}
\right].
\end{equation}
Note that all the relevant matrices have been carefully symmetrized.
Also note the important distinction between $C_0^2$ and $C^2$.  Now
define
\begin{equation}
c^2 = {1\over2}\tr[C^2],
\end{equation}
then
\begin{equation}
\left.{\d\omega^2\over\d k^2}\right|_{k\to0} = c^2 (1\pm \eta_2),
\end{equation}
with
\begin{equation}
\eta_2 
= \left\{ 
{ \tr[C^2]\tr[\Omega^2] + 2 \tr(\Omega^2\bar C_0^2 ) 
\over \tr[\Omega^2] \tr[C^2]}
\right\}
 = \left\{ 1 + {\tr(\Omega^2\bar C_0^2 ) \over \omega_{II}^2 \; c^2}
\right\}.
\end{equation}
Similarly, consider the second derivative: 
\begin{eqnarray}
{\d^2\omega^2\over\d(k^2)^2} &=&  {1\over2}\Bigg[\tr[H_s''(k^2)] 
\nonumber
\\
&&
\pm 
{
\tr[H_s(k^2)]\tr[H_s''(k^2)] + \tr[H_s'(k^2)]\tr[H_s'(k^2)]- 2\det''[H(k^2)] \over
\sqrt{ \tr[H_s(k^2)]^2- 4\det[H(k^2)] }
}
\nonumber
\\
&&
\mp 
{
( \tr[H_s(k^2)]\tr[H_s'(k^2)] - 2\det'[H(k^2)] )^2\over
(\tr[H_s(k^2)]^2- 4\det[H(k^2)] )^{3/2}
}
\Bigg],
\end{eqnarray}
whence
\begin{eqnarray}
\left.{\d^2\omega^2\over\d(k^2)^2}\right|_{k\to0} &=&  {1\over2}\Bigg[\tr[Z^2] 
\pm 
{
\tr[\Omega^2]\tr[Z^2] + \tr[C^2]^2- 2\det''[H(k^2)]_{k\to0} \over
\tr[\Omega^2]
}
\nonumber
\\
&&
\mp 
{
( \tr[\Omega^2]\tr[C^2] - 2\det'[H(k^2)]_{k\to0} )^2\over
\tr[\Omega^2]^3
}
\Bigg].
\end{eqnarray}
The last term above can be related to $\d\omega^2/\d k^2$, while the
determinant piece is evaluated using
\begin{eqnarray}
{\det}''[H(k^2)] &=& (2\det[C_0^2] ) \; \det[I + k^2 Y^2/2] 
\\
&&
+ (-\tr(\Omega^2\bar C_0^2)+ 2 k^2 \det[C_0^2] ) \;  (-\tr[\bar Y^2] +  k^2 \det[Y^2])/2
\nonumber
\\
&&
+  \det[\Omega^2+C_0^2 k^2] \; ( \det[Y^2]/2)
\nonumber
\\
&&
+   (-\tr(\Omega^2\bar C_0^2)+ 2 k^2 \det[C_0^2] )  \; (-\tr[\bar Y^2] +  k^2 \det[Y^2])/2.
\nonumber
\end{eqnarray}
Therefore
\begin{eqnarray}
{\det}''[H(k^2)]_{k\to0} &=& (2\det[C_0^2] ) 
\nonumber
\\
&&
+ (-\tr(\Omega^2\bar C_0^2) ) \;  (-\tr[\bar Y^2] )/2
+  \det[\Omega^2] \; (\det[Y^2])/2
\nonumber
\\
&&
+   (-\tr(\Omega^2\bar C_0^2))  \; (-\tr[\bar Y^2])/2.
\end{eqnarray}
That is, (recalling $\tr[\bar A] = - \tr[A]$),
\begin{equation}
{\det}''[H(k^2)]_{k\to0} = (2\det[C_0^2] ) 
-(\tr(\Omega^2\bar C_0^2) ) \;  (\tr[Y^2] ),
\end{equation}
or
\begin{equation}
{\det}''[H(k^2)]_{k\to0} = -\tr[C_0^2 \bar C_0^2]
- \tr[\Omega^2\bar C_0^2] \;  \tr[Y^2] .
\end{equation}
Now assembling all the pieces, a little algebra yields
\begin{eqnarray}
 \left.{\d^2\omega^2\over\d(k^2)^2}\right|_{k\to0} =  
&{\textstyle 1\over \textstyle 2}&\Bigg[
\tr[Z^2]  \pm \tr[Z^2] 
\pm 2 \frac{\tr[\Omega^2\bar{C}^2_0]}{\tr[\Omega^2]}\tr[Y^2]
\pm {\tr[C^2]^2- 4\det[C^2_0] \over \tr[\Omega^2] }\nonumber\\
&\mp& 
{\tr[C^2]^2\over\tr[\Omega^2]} \eta_2^2
\Bigg].\label{eq: varpi4}
\end{eqnarray}

With the above formula we have completed our derivation of the
lowest-order terms of the generic dispersion relation of a coupled
2-BEC \index{2-BEC~system} system --- including the terms introduced
by the quantum potential at high wavenumber --- up to terms of order
$k^4$.  From the above formula it is clear that we do not generically
have Lorentz \index{Lorentz~invariance} invariance in this system:
Lorentz \index{Lorentz~invariance!violation} violations arise both due
to mode-mixing interactions (an effect which can persist in the
hydrodynamic \index{hydrodynamic~limit} limit where $Z\to0$ and
$Y\to0$) and to the presence of the quantum potential (signaled by
$Z\neq0$ and $Y\neq0$).  While the mode-mixing effects are relevant at
all energies the latter effect characterizes the discrete structure of
the effective spacetime at high energies. It is in this sense that the
quantum potential determines the analogue of quantum gravity effects
in our 2-BEC \index{2-BEC~system} system.
%

%+++++++++++++++++++++++++++++++++++++++++++++++++++++++++++++++++
\subsection{The relevance for quantum gravity phenomenology\label{sec:qgp}}
%+++++++++++++++++++++++++++++++++++++++++++++++++++++++++++++++++

Following this physical insight we can now easily identify a regime
that is potentially relevant for simulating the typical ans\"atze of
quantum gravity\index{quantum~gravity!phenomenology} phenomenology.
We demand that any violation of Lorentz
\index{Lorentz~invariance!violation} invariance present should be due
to the microscopic structure of the effective spacetime. This implies
that one has to tune the system in order to cancel exactly all those
violations of Lorentz \index{Lorentz~invariance!violation} invariance
which are solely due to mode-mixing interactions in the hydrodynamic
\index{hydrodynamic~limit} limit.

We basically follow the guiding idea that a good analogue of
quantum-gravity-induced Lorentz \index{Lorentz~invariance!violation}
violations should be characterized only by the ultraviolet physics of
the effective spacetime. In the system at hand the ultraviolet physics
is indeed characterized by the quantum potential, whereas possible
violations of the Lorentz \index{Lorentz~invariance!violation}
invariance in the hydrodynamical limit are low energy effects, even
though they have their origin in the microscopic interactions. We
therefore start by investigating the scenario in which the system is
tuned in such a way that no violations of Lorentz
\index{Lorentz~invariance} invariance are present in the hydrodynamic
\index{hydrodynamic~limit} limit. This leads us to again enforce the
conditions $C1$ and $C2$ which corresponded to ``mono-metricity" in
the hydrodynamic \index{hydrodynamic~limit} limit.

In this case (\ref{eq: varpi2}) and  (\ref{eq: varpi4}) take respectively the form
\begin{eqnarray}
\left.{\d\omega^2\over\d k^2}\right|_{k\to0} &=& 
{1\over2}\left[\tr[C_0^2] +(1\pm1)\,\tr[\Delta C^2]\right]
= c_0^2 + {1\pm1\over2}\tr[\Delta C^2],\label{eq:varpi2b}
\end{eqnarray}
and
\begin{eqnarray}
\left.{\d^2\omega^2\over\d(k^2)^2}\right|_{k\to0} &=&
{\tr[Z^2]  \pm \tr[Z^2] \over 2}\mp\tr[C^2_0]\tr[Y^2]
\nonumber
\\
&&
\pm{1\over 2} {\tr[\Delta C^2]^2+2\tr[C^2_0]\tr[\Delta C^2] \over \tr[\Omega^2] }
\mp {1\over 2}
{
\tr[\Delta C^2]^2\over
\tr[\Omega^2]
}
\nonumber
\\
&=&
{\tr[Z^2]  \pm \tr[Z^2] \over 2}\pm\tr[C^2_0]\left(-\tr[Y^2]+
{\tr[\Delta C^2] \over \tr[\Omega^2] }\right).
\label{eq:varpi4b}
\end{eqnarray}

Recall (see section \ref{S:C1C2}) that the first of the physical
conditions $C1$ is equivalent to the statement that the $2\times2$
matrix $C_0^2$ has two identical eigenvalues. But since $C_0^2$ is
symmetric this then implies $C_0^2 = c_0^2 \; \mathbf{I}$, in which
case the second condition is automatically satisfied.  This also leads
to the useful facts
\begin{eqnarray}
&&\tilde{U}_{\A\B}=0 \quad \Longrightarrow \quad \lambda = -2 \sqrt{\rho_{\A0}\;\rho_{\B0}} \; U_{\A\B}
\label{eq:lambdaUAB} ;\\
&& c_0^2 ={\tilde U_{\A\A} \;\rho_{\A0}\over m_\A} =  {\tilde
  U_{\B\B} \;\rho_{\B0}\over m_\B}.
    \label{eq:c0ft2}
\end{eqnarray}
Now that we have the fine tuning condition for the laser coupling we
can compute the magnitude of the effective mass of the massive
phonon\index{phonon} and determine the values of the Lorentz
\index{Lorentz~invariance!violation} violation coefficients. In
particular we shall start checking that this regime allows for a real
positive effective mass as needed for a suitable analogue model of
quantum gravity\index{quantum~gravity!phenomenology} phenomenology.

%~~~~~~~~~~~~~~~~~~~~~~~~~~~~~~~~~~~~~~~~~~~~~~~~~~~~
\subsubsection{Effective mass}
%~~~~~~~~~~~~~~~~~~~~~~~~~~~~~~~~~~~~~~~~~~~~~~~~~~~~
Remember that  the definition of $m_{II}$ reads 
\begin{equation}
m_{II}^2 = \hbar^2 \omega_{II}^2/c_0^4.
\label{eq:m2-c1}
\end{equation}
Using equation~(\ref{eq:lambdaUAB}) and equation~(\ref{eq:c0ft2})
we can rewrite $c^2_0$ in the following form
\begin{equation}
c_0^2 = [m_\B \rho_{\A0} U_{\A\A} + m_\A \rho_{\B0} U_{\B\B} 
         + U_{\A\B} (\rho_{\A0} m_\A + \rho_{\B0} m_\B)] /
         (2 m_\A m_\B).
         \label{eq:c0av}
\end{equation}
Similarly equation~(\ref{eq:lambdaUAB}) and equation~(\ref{eq:c0ft2})
when inserted in equation (\ref{E:m2}) give
\begin{equation}
\omega_{II}^2 = \frac{4 U_{\A\B} (\rho_{\A0} m_\B + \rho_{\B0} m_\A) c_0^2}{ \hbar^2}.
\label{eq:om2}
\end{equation}
We can now estimate $m_{II}$ by simply inserting the above expressions
in equation~(\ref{eq:m2-c1}) so that
\begin{equation}
m_{II}^2 = {
8 U_{\A\B} (\rho_{\A0} m_\A+\rho_{\B0} m_\B) m_\A m_\B 
\over
          [m_\B \rho_{\A0} U_{\A\A} + m_\A \rho_{\B0} U_{\B\B} 
         + U_{\A\B} (\rho_{\A0} m_\A + \rho_{\B0} m_\B)]
}.
\end{equation}
This formula is still a little clumsy but a great deal can be
understood by doing the physically reasonable approximation $m_\A
\approx m_\B=m$ and $\rho_\A \approx \rho_\B$. In fact in this case one
obtains
\begin{equation}
m_{II}^2 \approx m^2  \; {
8 U_{\A\B}
\over [U_{\A\A}+2U_{\A\B}+U_{\B\B}]
}.
\end{equation}
This formula now shows clearly that, as long as the mixing term
$U_{\A\B}$ is small compared to the ``direct" scattering
$U_{\A\A}+U_{\B\B}$, the mass of the heavy phonon\index{phonon} will
be ``small" compared to the mass of the atoms.  Though experimental
realizability of the system is not the primary focus of the current
article, we point out that there is no obstruction in principle to
tuning a 2-BEC \index{2-BEC~system} system into a regime where
$|U_{\A\B}| \ll |U_{\A\A}+U_{\B\B}|$.  For the purposes of this paper
it is sufficient that a small effective phonon\index{phonon} mass
(small compared to the atomic masses which set the analogue quantum
gravity scale) is obtainable for some arrangement of the microscopic
parameters.
We can now look separately at the coefficients of the quadratic and
quartic Lorentz \index{Lorentz~invariance!violation} violations and
then compare their relative strength in order to see if a situation
like that envisaged by discussions of the naturalness problem is
actually realized.

%~~~~~~~~~~~~~~~~~~~~~~~~~~~~~~~~~~~~~~~~~~~~~~~~~~~~
\subsubsection{Coefficient of the quadratic deviation}
%~~~~~~~~~~~~~~~~~~~~~~~~~~~~~~~~~~~~~~~~~~~~~~~~~~~~
One can easily see from (\ref{eq:varpi2b}) that the $\eta_2$
coefficients for this case take the form
\begin{eqnarray}
\eta_{2,I} &=& 0;\\
\eta_{2,II}\; c_0^2  &=&\tr[ \Delta C^2]=\tr[X^{1/2} \Lambda X^{1/2}]=\tr[X\Lambda]\nonumber
\\
&=& -\frac{1}{2}\frac{\lambda}{m_\A m_\B}\left( 
\frac{m_\A\rho_{\A0}+m_\B\rho_{\B0}}
{\sqrt{\rho_{\A0}\rho_{\B0}}} \right).
\end{eqnarray}

%%%%%%%
So if we insert the fine tuning condition for $\lambda$,
equation~(\ref{eq:lambdaUAB}), we get
\begin{eqnarray}
\eta_{2,II} &=& 
=\frac{U_{\A\B}\left( 
m_\A\rho_{\A0}+m_\B\rho_{\B0}\right)
}{m_\A m_\B c_0^2}.
\label{E:eta_2II}
\end{eqnarray}
Remarkably we can now cast this coefficient in a much more suggestive
form by expressing the coupling $U_{\A\B}$ in terms of the mass of the
massive quasi-particle\index{quasi-particle} $m_{II}^2$.  In order to do this we start from
equation (\ref{eq:om2}) and note that it enables us to express
$U_{\A\B}$ in (\ref{E:eta_2II}) in terms of $\omega_{II}^2$, thereby
obtaining
\begin{equation}
\eta_{2,II} =\frac{\hbar^2}{4 c^4_0} \; 
\frac{\rho_{\A0} m_\A + \rho_{\B0} m_\B}{\rho_{\A0} m_\B + \rho_{\B0} m_\A} \;
\frac{\omega_{II}^2 }{m_\A m_\B}.
\end{equation}
Now it is easy to see that
\begin{equation}
\frac{\rho_{\A0} m_\A + \rho_{\B0} m_\B}{\rho_{\A0} m_\B + \rho_{\B0} m_\A}
\approx \mathcal{O} (1),
\end{equation}
and that this factor is identically unity if either $m_\A = m_\B$ or $\rho_{\A0} = \rho_{\B0}$.
All together we are left with
\begin{equation} 
\eta_{2,II}  = \bar{\eta} \left(
\frac{m_{II} }{\sqrt{m_\A m_\B}} \right)^2, 
\label{eq:eta2qfin}
\end{equation}
where $\bar{\eta}$ is a dimensionless coefficient of order unity.

The product in the denominator of the above expression can be
interpreted as the geometric mean of the fundamental bosons masses
$m_\A$ and $m_\B$. These are mass scales associated with the
microphysics of the condensate --- in analogy with our experience with
a 1-BEC \index{1-BEC~system} system where the ``quantum gravity
scale'' is set by the mass of the BEC atoms. It is then natural to
define an analogue of the scale of the breakdown of Lorentz
\index{Lorentz~invariance!violation} invariance as $M_{\rm
  eff}=\sqrt{m_\A m_\B}$.  (Indeed this ``analogue Lorentz breaking
scale'' will typically do double duty as an ``analogue Planck
\index{Planck~scale} mass".)

Using this physical insight it should be clear that
equation~(\ref{eq:eta2qfin}) effectively says
\begin{equation}
\eta_{2,II}\approx\left(\frac{m_{II}}{M_{\rm eff}}\right)^2,
\label{eq:eta2fin}
\end{equation}
which, given that $m_I=0$, we are naturally lead to generalize to
\begin{equation}
\eta_{2,X}\approx\left(\frac{m_X}{M_{\rm eff}}\right)^2= 
\left( {\hbox{mass scale of quasiparticle}
\over
\hbox{effective Planck scale}}\right)^2; \qquad X=I,II.
\label{eta2_final}
\end{equation}
The above relation is exactly the sort of dimensionless ratio
$(\mu/M)^\sigma$ that has been very often \emph{conjectured} in the
literature on quantum gravity\index{quantum~gravity!phenomenology}
phenomenology in order to explain the strong observational constraints
on Lorentz \index{Lorentz~invariance!violation} violations at the
lowest orders. (See earlier discussion.)
Does this now imply that this particular regime of our 2-BEC
\index{2-BEC~system} system will also show an analogue version of the
naturalness problem? In order to answer this question we need to find
the dimensionless coefficient for the quartic deviations, $\eta_4$,
and check if it will or won't itself be suppressed by some power of
the small ratio $m_{II}/M_{\rm eff}$.

%~~~~~~~~~~~~~~~~~~~~~~~~~~~~~~~~~~~~~~~~~~~~~~~~~~~~
\subsubsection{Coefficients of the quartic deviation}
%~~~~~~~~~~~~~~~~~~~~~~~~~~~~~~~~~~~~~~~~~~~~~~~~~~~~
Let us now consider the coefficients of the quartic term presented in
equation~(\ref{eq:varpi4b}).  For the various terms appearing in
(\ref{eq:varpi4b}) we get
\begin{equation}
\tr[Z^2]=2\tr[DX]=
\frac{\hbar^2}{2}\left(\frac{m^2_\A+m^2_\B}{m^2_\A m^2_\B}\right);
\end{equation}
\begin{equation}
\tr[\Delta C^2]=\tr[X \Lambda]=
-\frac{\lambda}{2}\frac{m_\A\rho_{\A0}+m_\B\rho_{\B0}}
{m_\A m_\B\sqrt{\rho_{\A0}\rho_{\B0}}}
=
U_{\A\B}\frac{m_\A\rho_{\A0}+m_\B\rho_{\B0}}{m_\A m_\B};
\end{equation}
\begin{equation}
\tr[Y^2]=2\tr[X\Xi^{-1}]=\frac{\hbar^2}{2}
\frac{\rho_{\A0}m_{\A}\tilde{U}_{\A\A}+\rho_{\B0}m_{\B}\tilde{U}_{\B\B}}
{\rho_{\A0}m_{\A}\rho_{\B0} m_{\B} \tilde{U}_{\A\A}\tilde{U}_{\B\B}};
\end{equation}
where in the last expression we have used the fact that in the current
scenario $\tilde{U}_{\A\B}=0$.
Now by definition 
\begin{equation}
\eta_4 = 
{1\over2} ({M_\mathrm{eff}^2}/{\hbar^2}) \left[{\d^2\omega^2\over (\d k^2)^2}\right]_{k=0}
\end{equation}
 is the dimensionless coefficient in front of the $k^4$. So
\begin{eqnarray}
\eta_4 &=& 
\frac{M_\mathrm{eff}^2}{2\hbar^2} \left[ {\tr[Z^2]  \pm 
\tr[Z^2] \over 2}\pm\tr[C^2_0]\left(-{\tr[Y^2]\over2}+
{\tr[\Delta C^2] \over \tr[\Omega^2] }\right)\right]\\
&=& 
\frac{ M_\mathrm{eff}^2\; c_0^2}{\hbar^2} 
\left[ 
{\tr[Z^2]  \pm \tr[Z^2] \over 2\tr[C^2_0]}\pm\left(-{\tr[Y^2]\over2}+
{\tr[\Delta C^2] \over \tr[\Omega^2] }\right)\right].
\end{eqnarray}
Whence 
\begin{eqnarray}
\eta_{4,I} &=& \frac{ M_\mathrm{eff}^2\;c_0^2}{\hbar^2}
\left[ {\tr[Z^2] \over \tr[C^2_0]}+\left(-{\tr[Y^2]\over2}+
{\tr[\Delta C^2] \over \tr[\Omega^2] }\right)\right];
\\
\eta_{4,II} &=& \frac{ M_\mathrm{eff}^2\;c_0^2}{\hbar^2}
\left[ \left({\tr[Y^2]\over2}-
{\tr[\Delta C^2] \over \tr[\Omega^2] }\right)\right].
\end{eqnarray}
Let us compute the two relevant terms separately:
\begin{eqnarray}
{\tr[Z^2] \over \tr[C^2_0]} &=&
\frac{\hbar^2}{4c_0^2}\left(\frac{m^2_\A+m^2_\B}{m^2_\A m^2_\B}\right)=
\frac{\hbar^2}{4 c_0^2 M_\mathrm{eff}^2}
\left(\frac{m^2_\A+m^2_\B}{m_\A m_\B}\right);
\end{eqnarray}
\begin{eqnarray}
-\tr[Y^2]/2+
{\tr[\Delta C^2] \over \tr[\Omega^2] } &=& 
-\frac{\hbar^2}{4 M_\mathrm{eff}^2}
\left[
\frac{\rho_{\A0}m_{\A}\tilde{U}_{\A\A}^2+\rho_{\B0}m_{\B}\tilde{U}_{\B\B}^2}
{\rho_{\A0}\rho_{\B0}\tilde{U}_{\A\A}\tilde{U}_{\B\B}
\left(\tilde{U}_{\A\A}+\tilde{U}_{\B\B}\right)}
\right]
\nonumber\\
&=& 
-\frac{\hbar^2}{4 M_\mathrm{eff}^2\;c_0^2}
\left[
\frac{m^2_{\A}\tilde{U}_{\A\A}+m^2_{\B}\tilde{U}_{\B\B}}
{m_{\A}m_{\B}\left(\tilde{U}_{\A\A}+\tilde{U}_{\B\B}\right)}
\right];
\end{eqnarray}
where we have used $\rho_{X0}\tilde{U}_{XX}=m_{X}c^2_0$ for $X=A,B$ as
in equation~(\ref{eq:c0ft2}).  Note that the quantity in square
brackets in the last line is dimensionless.  So in the end:
\begin{eqnarray}
 \eta_{4,I} &=&  \frac{1}{4}\left[
\left(\frac{m^2_\A+m^2_\B}{m_\A m_\B}\right)
-\frac{m^2_{\A}\tilde{U}_{\A\A}+m^2_{\B}\tilde{U}_{\B\B}}
{m_{\A}m_{\B}\left(\tilde{U}_{\A\A}+\tilde{U}_{\B\B}\right)}
\right]\\
&=&
\frac{1}{4}
\left[
\frac{m^2_{\A}\tilde{U}_{\B\B}+m^2_{\B}\tilde{U}_{\A\A}}
{m_{\A}m_{\B}\left(\tilde{U}_{\A\A}+\tilde{U}_{\B\B}\right)}\right];
\label{eq:eta4I}\\
 \eta_{4,II} &=& \frac{1}{4}
\left[
\frac{m^2_{\A}\tilde{U}_{\A\A}+m^2_{\B}\tilde{U}_{\B\B}}
{m_{\A}m_{\B}\left(\tilde{U}_{\A\A}+\tilde{U}_{\B\B}\right)}\right].
\label{eq:eta4II}
\end{eqnarray}
\emph{Note:} In the special case $m_\A=m_\B$ we recover identical
quartic deviations $\eta_{4,I}=\eta_{4,II}=1/4$, indicating in this
special situation a ``universal'' deviation from Lorentz
\index{Lorentz~invariance} invariance.  Indeed we also obtain
$\eta_{4,I}=\eta_{4,II}$ if we demand $\tilde{U}_{\A\A} =
\tilde{U}_{\B\B}$, even without fixing $m_\A=m_\B$.

Thus in the analogue spacetime\index{analogue~spacetime} we have
developed the issue of {\em universality} is fundamentally related to
the complexity of the underlying microscopic system. As long as we
keep the two atomic masses $m_\A$ and $m_\B$ distinct we generically
have distinct $\eta_4$ coefficients (and the $\eta_2$ coefficients are
unequal even in the case $m_\A=m_\B$). However we can easily recover
identical $\eta_4$ coefficients, for instance, as soon as we impose
identical microphysics for the two BEC systems we couple.

%~~~~~~~~~~~~~~~~~~~~~~~~~~~~~~~~~~~~~~~~~~~~~~~~~~~~
\subsubsection{Avoidance of the naturalness problem}
%~~~~~~~~~~~~~~~~~~~~~~~~~~~~~~~~~~~~~~~~~~~~~~~~~~~~
We can now ask ourselves if there is, or is not, a naturalness problem
present in our system. Are the dimensionless coefficients
$\eta_{4,I/II}$ suppressed below their naive values by some small
ratio involving $M_{\rm eff}=\sqrt{m_\A m_\B}$~? Or are these ratios
unsuppressed?  Indeed at first sight it might seem that further
suppression is the case, since the square of the ``effective Planck
\index{Planck~scale} scale" seems to appear in the denominator of both
the coefficients (\ref{eq:eta4I}) and (\ref{eq:eta4II}). However, the
squares of the atomic masses also appear in the numerator, rendering
both coefficients of order unity.

It is perhaps easier to see this once the dependence of
(\ref{eq:eta4I}) and (\ref{eq:eta4II}) on the effective coupling
$\tilde{U}$ is removed. We again use the substitution
$\tilde{U}_{XX}=m_{X}c^2_0/\rho_{X0}$ for $X=\A,\B$, so obtaining:
\begin{eqnarray}
\eta_{4,I} &=&  \frac{1}{4}
\left[
\frac{m_{\A}\rho_{\A0}+m_{\B}\rho_{\B0}}{m_{\A}\rho_{\B0}+m_{\B}\rho_{\A0}}
\right];
\label{eq:eta4Ifin}\\
&&\nonumber\\
\eta_{4,II} &=& \frac{1}{4}
\left[
\frac{m_{\A}^3 \rho_{\B0}+m_{\B}^3 \rho_{\A0}} 
{
m_\A m_\B  \;( m_{\A}\rho_{\B0}+m_{\B}\rho_{\A0})}
\right].
\label{eq:eta4IIfin}
\end{eqnarray}
{From} these expressions is clear that the $\eta_{4,I/II}$ coefficients
are actually of order unity.

That is, if our system is set up so that $m_{II}\ll m_{\A/\B}$ --- which
we have seen in this scenario is equivalent to requiring $U_{\A\B}\ll
U_{\A\A/\B\B}$ --- no naturalness problem arises as for $p > m_{II}\;c_0$
the higher-order, energy-dependent Lorentz-violating terms ($n\geq 4$)
will indeed dominate over the quadratic Lorentz-violating term.

It is quite remarkable that the quadratic coefficients
(\ref{eta2_final}) are {\em exactly} of the form postulated in several
works on non-renormalizable \EFT with Lorentz
\index{Lorentz~invariance!violation} invariance violations (see
e.g.~\cite{jlm-ann}).  They are indeed the squared ratio of the
particle mass to the scale of Lorentz
\index{Lorentz~invariance!violation} violation.  Moreover we can see
from (\ref{eq:eta4I}) and (\ref{eq:eta4II}) that there is no further
suppression --- after having pulled out a factor $(\hbar /
M_{\mathrm{Lorentz~violation}})^2$ --- for the quartic coefficients
$\eta_{4,\mathrm{I/|I}}$. These coefficients are of order one and
generically non-universal, (though if desired they can be forced to be
universal by additional and specific fine tuning).

The suppression of $\eta_2$, combined with the \emph{non-suppression}
of $\eta_4$, is precisely the statement that the ``naturalness
problem'' does not arise in the current model. We stress this is not a
``tree level'' result as the dispersion relation was computed directly
from the fundamental Hamiltonian and was not derived via any \EFT
reasoning.  Moreover avoidance of the naturalness problem is not
directly related to the tuning of our system to reproduce special relativity in the
hydrodynamic \index{hydrodynamic~limit} limit. In fact our conditions
for recovering special relativity at low energies do not \emph{a priori} fix the the
$\eta_2$ coefficient, as its strength after the ``fine tuning" could
still be large (even of order one) if the typical mass scale of the
massive phonon\index{phonon} is not well below the atomic mass scale.
Instead the smallness of $\eta_2$ is directly related to the
mass-generating \index{mass-generating~mechanism} mechanism.

The key question is now: Why does our model escape the naive
predictions of dominant lowest-dimension Lorentz
\index{Lorentz~invariance!violation} violations?
(In fact in our model for any $p\gg m_{II}$ the $k^4$ Lorentz
\index{Lorentz~invariance!violation} violating term dominates over the
order $k^2$ one.)  We here propose a nice interpretation in terms of
``emergent symmetry'':\index{emergent~symmetry} Non-zero $\lambda$
\emph{simultaneously} produces a non-zero mass for one of the
phonons,\index{phonon} \emph{and} a corresponding non-zero Lorentz
\index{Lorentz~invariance!violation} violation at order $k^2$.
(Single BEC systems have only $k^4$ Lorentz
\index{Lorentz~invariance!violation} violations as described by the
Bogoliubov dispersion relation.)  Let us now drive $\lambda\to 0$, but
keep the conditions $C1$ and $C2$ valid at each stage. (This also
requires $U_{\A\B}\to 0$.) One gets an
\EFT\index{effective~field~theory} which at low energies describes two
non-interacting phonons\index{phonon} propagating on a common
background. (In fact $\eta_2\to0$ and $c_I=c_{II}=c_0$.) This system
possesses a $SO(2)$ symmetry.  Non-zero laser coupling $\lambda$
softly breaks this $SO(2)$, the mass degeneracy, and low-energy
Lorentz \index{Lorentz~invariance} invariance. Such soft Lorentz
\index{Lorentz~invariance!violation} violation is then characterized
(as usual in \EFT)\index{effective~field~theory} by the ratio of the
scale of the symmetry breaking $m_{II}$, and that of the scale
originating the Lorentz \index{Lorentz~invariance!violation} violation
in first place $M_{\rm Lorentz~violation}$. We stress that the $SO(2)$ symmetry is
an ``emergent\index{emergent~symmetry} symmetry'' as it is not
preserved beyond the hydrodynamic \index{hydrodynamic~limit} limit:
the $\eta_4$ coefficients are in general different if $m_\A\neq m_\B$,
so $SO(2)$ is generically broken at high energies.  Nevertheless this
is enough for the protection of the {\em lowest}-order Lorentz
\index{Lorentz~invariance!violation} violating operators.  The lesson
to be drawn is that emergent symmetries are sufficient to minimize the
amount of Lorentz \index{Lorentz~invariance!violation} violation in
the lowest-dimension operators of the
\EFT.\index{effective~field~theory} In this regard, it is intriguing to
realise that an interpretation of SUSY as an accidental symmetry has
indeed been considered in recent times~\cite{Luty}, and that this is
done at the cost of renouncing attempts to solve the hierarchy problem
in the standard way. It might be that in this sense the smallness of
the particle physics mass scales with respect to the Planck scale
could be directly related to smallness of Lorentz violations in
renormalizable operators of the low-energy effective field theory we
live in. We hope to further investigate these issues in future work.
%

%%%%%%%%%%%%%%%%%%%%%%%%%%%%%%%%%%%%%%%%%%%%%%%%%%%

\section{Outlook, summary and discussion}                                     
               
%%%%%%%%%%%%%%%%%%%%%%%%%%%%%%%%%%%%%%%%%%%%%%%%%%%

So where can (and should) we go from here? If 2-component BECs provide
such a rich mathematical and physical structure, are 3-component BECs,
or general multi-component BECs even better? That depends on what you
are trying to do:
\begin{itemize}
\item If one wishes to actually \emph{build} such an
  analogue\index{analogue~spacetime} spacetime in the laboratory, and
  perform actual experiments, then iteration through 1-BEC
  \index{1-BEC~system} and 2-BEC \index{2-BEC~system} systems seems
  the most promising route in terms of our technological capabilities.
\item For $n$-component BECs we sketch the situation in figure
  \ref{fig:5_{comp_system}}. The key point is that due to overall
  translation invariance one again expects to find one massless
  quasi-particle, with now $n-1$ distinct massive modes. Unfortunately
  the matrix algebra is now considerably messier --- not intrinsically
  difficult (after all we are only dealing with $n\times n$ matrices
  in field space) --- but extremely tedious. Physical insight remains
  largely intact, but (except in some specific particularly simple
  cases), computations rapidly become lost in a morass of technical
  detail.
\item However, if one wishes to draw general theoretical lessons from
  the analogue spacetime\index{analogue~spacetime} programme, then
  multi-component systems are definitely the preferred route ---
  though in this case it is probably better to be even more abstract,
  and to go beyond the specific details of BEC-based systems to deal
  with general hyperbolic \index{hyperbolic~system} systems of PDEs.
\item In appendix A we have sketched some of the key features of the
  pseudo--Finsler \index{pseudo--Finsler~spacetime} spacetimes that
  naturally emerge from considering the leading symbol of a hyperbolic
  \index{hyperbolic~system} system of PDEs. While it is clear that
  much more could be done based on this, and on extending the field
  theory ``normal modes'' of~\cite{normal,normal2}, such an analysis
  would very much move outside the scope of the COSLAB \index{COSLAB}
  programme.
\end{itemize}
In short the 2-BEC \index{2-BEC~system} system is a good compromise
between a system complex enough to exhibit a mass-generating
\index{mass-generating~mechanism} mechanism, and still simple enough
to be technologically tractable, with good prospects for laboratory
realization of this system in the not too distant future.
%_figure__figure__figure__figure__figure__figure__figure__figure__figure__figure__figure__figure_
\begin{figure}[!htb]
\centering
\includegraphics[height=1.9cm]{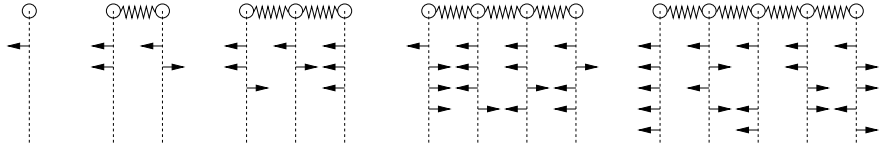}
\caption{The figure captures the key features of possible
  eigenmodes\index{eigenmodes} for a small perturbation (circles) in a
  1 (left side), 2, 3, 4, and 5-component (right side) BEC. In a
  1-component system only one kind of perturbation is allowed, which
  corresponds to a massless particle propagating through an effective
  curved spacetime, while in the 2-component case two different kinds
  of mode appear, the one in-phase (massless particle) and one in
  anti-phase (massive particle).  For a three-component system we
  again expect to find one massless particle, when all perturbations
  are in phase, and now in addition to that two massive particles.}
\label{fig:5_{comp_system}}       
\end{figure}
%_figure__figure__figure__figure__figure__figure__figure__figure__figure__figure__figure__figure_
%

The key features we have emphasised in this chapter have been:
\begin{itemize}
\item A general analysis of the 2-BEC \index{2-BEC~system} system to
  see how perturbations on a 2-BEC \index{2-BEC~system} background
  lead to a system of coupled wave \index{wave~equation} equations.
\item Extraction of the geometric notion of pseudo--Finsler
  \index{pseudo--Finsler~spacetime} spacetime from this
  wave\index{wave~equation} equation, coupled with an analysis of how
  to specialize pseudo--Finsler \index{pseudo--Finsler~spacetime}
  geometry first to a bi-metric \index{bi-metric~spacetime} Lorentzian
  geometry and finally to the usual mono-metric
  \index{mono-metric~spacetime} Lorentzian geometry of most direct
  interest in general relativity and cosmology.
\item The mass-generating \index{mass-generating~mechanism} mechanism
  we have identified in suitably coupled 2-component BECs is an
  essential step in making this analogue
  spacetime\index{analogue~spacetime} more realistic; whatever one's
  views on the ultimate theory of ``quantum
  gravity'',\index{quantum~gravity} any realistic low-energy
  phenomenology\index{quantum~gravity!phenomenology} must contain
  \emph{some} mass-generating \index{mass-generating~mechanism}
  mechanism.
\item Use of the ``quantum \index{quantum~pressure} pressure'' term in
  the 2-BEC \index{2-BEC~system} system to mimic the sort of Lorentz
  \index{Lorentz~invariance!violation} violating physics that (based
  on the relatively young field of ``quantum
  gravity\index{quantum~gravity!phenomenology} phenomenology'') is
  widely expected to occur at or near the Planck
  scale.\index{Planck~scale}
\item Intriguingly, we have seen that in our specific model the
  mass-generating \index{mass-generating~mechanism} mechanism
  interacts with the Lorentz \index{Lorentz~invariance!violation}
  violating mechanism, naturally leading to a situation where the
  Lorentz \index{Lorentz~invariance!violation} violations are
  suppressed by powers of the quasi-particle\index{quasi-particle}
  mass scale divided by the analogue of the Planck
  scale.\index{Planck~scale}
\end{itemize}

In summary, while we do not personally believe that the real universe
\emph{is} an analogue spacetime,\index{analogue~spacetime} we are
certainly intrigued by the fact that so much of what is normally
viewed as being specific to general relativity and/or particle physics
can be placed in this much wider context. We should also be forthright
about the key weakness of analogue models as they currently stand: As
we have seen, obtaining an analogue
spacetime\index{analogue~spacetime} geometry (including spacetime
curvature) is straightforward --- but what is not straightforward is
obtaining the Einstein equations.\index{Einstein~equations} The
analogue models are currently analogue models of quantum field theory
on curved spacetime, but not (yet?) true analogue models of Einstein
gravity.\index{Einstein~gravity} Despite this limitation, what can be
achieved through the analogue spacetime\index{analogue~spacetime}
programme is quite impressive, and we expect interest in this field,
both theoretical and hopefully experimental, to continue unabated.

%%%%%%%%%%%%%%%%%%%%%%%%%%%%%%%%%%%%%%%%%%%%%%%%%%%
%%%%%%%%%%%%%%%%%%%%%%%%%%%%%%%%%%%%%%%%%%%%%%%%%%%
%%%%%%%%%%%%%%%%%%%%%%%%%%%%%%%%%%%%%%%%%%%%%%%%%%%
%%%%%%%%%%%%%%%%%%%%%%%%%%%%%%%%%%%%%%%%%%%%%%%%%%%
%%%%%%%%%%%%%%%%%%%%%%%%%%%%%%%%%%%%%%%%%%%%%%%%%%%
%%%%%%%%%%%%%%%%%%%%%%%%%%%%%%%%%%%%%%%%%%%%%%%%%%%
%%%%%%%%%%%%%%%%%%%%%%%%%%%%%%%%%%%%%%%%%%%%%%%%%%%
%%%%%%%%%%%%%%%%%%%%%%%%%%%%%%%%%%%%%%%%%%%%%%%%%%%
\appendix
%%%%%%%%%%%%%%%%%%%%%%%%%%%%%%%%%%%%%%%%%%%%%%%%%%%

\section{Finsler and co--Finsler geometries}  

%%%%%%%%%%%%%%%%%%%%%%%%%%%%%%%%%%%%%%%%%%%%%%%%%%%

Finsler \index{Finsler!geometry} geometries are sufficiently unusual
that a brief discussion is in order --- especially in view of the fact
that the needs of the physics community are often somewhat at odds
with what the mathematical community might view as the most important
issues.  Below are some elementary results, where we emphasise that
for the time being we are working with ordinary ``Euclidean
signature'' Finsler \index{Finsler!geometry} geometry. For general
references, see~\cite{Finsler-general-references}.

%+++++++++++++++++++++++++++++++++++++++++++++++++++++++++++++++++
\subsection{Basics}
%+++++++++++++++++++++++++++++++++++++++++++++++++++++++++++++++++

\noindent\emph{Euler theorem:} If $H(z)$ is homogeneous of degree $n$ then
\begin{equation}
z^i\; {\partial H(z)\over\partial z^i} = n \; H(z).
\end{equation}

\noindent\emph{Finsler \index{Finsler!function} function:} Defined on
the ``slit tangent bundle'' $T_{\neq0}(M)$ such that $F:T_{\neq0}(M)\to[0,+\infty)$
where
\begin{equation}
F(x,t): \quad F(x,\lambda  t) = \lambda \; F(x, t),
\end{equation}
and
\begin{equation}
T_{\neq0}(M) = \bigcup_{x\in M} \left[ T_x-\{\vec 0\} \right].
\end{equation}
That is, the Finsler function is a defined only for nonzero tangent
vectors $t \in [T_x-\{\vec 0\}]$, and for any fixed direction is
linear in the size of the vector.

\noindent\emph{Finsler \index{Finsler!distance} distance:} 
\begin{equation}
d_\gamma(x,y) = \int_x^y F(x(\tau),dx/d\tau) \; d\tau; \qquad \tau=\;\hbox{arbitrary parameter}.
\end{equation}

\noindent\emph{Finsler \index{Finsler!metric} metric:} 
\begin{equation}
g_{ij}(x,t) = {1\over2} {\partial^2 [F^2(x,t)]\over\partial t^i\;\partial t^j}.
\end{equation}

\noindent
The first slightly unusual item is the introduction of co--Finsler
\index{co--Finsler~geometry} structure:
\\
\emph{co--Finsler function:} Define a co--Finsler
\index{co--Finsler~geometry} structure on the cotangent bundle by
Legendre transformation of $F^2(x,t)$. That is:
\begin{equation}
G^2(x,p) = t^j(p) \; p_j - F^2(x,t(p))
\end{equation}
where $t(p)$ is defined by the Legendre transformation condition
\begin{equation}
{\partial [F^2]\over\partial t^j}(x,t) = p_j.
\end{equation}
Note
\begin{equation}
{\partial p_j\over\partial t^k} = {\partial [F^2]\over \partial t^j\; \partial t^k} = 2 g_{jk}(x,t),
\end{equation}
which is why we demand the Finsler \index{Finsler!metric} metric be
nonsingular.

\noindent\emph{Lemma:} $G(x,p)$ defined in this way is homogeneous of degree 1.
\\
\emph{Proof:} Note
\begin{equation}
z^i\; {\partial H(z)\over\partial z^i} = n \; H(z)
\end{equation}
implies
\begin{equation}
z^i\; {\partial \over\partial z^i} \left[{\partial^m\over(\partial z)^m} H(z)\right]  
= (n-m) \;\left[{\partial^m\over(\partial z)^m} H(z)\right].
\end{equation}
In particular:
\begin{itemize}
\item $F^2$ is homogeneous of degree 2.
\item $g_{ij}$ is homogeneous of degree 0.
\item $\partial[F^2]/\partial t$ is homogeneous of degree 1.
\item Therefore $p(t)$  is homogeneous of degree 1 \\
  \emph{and} $t(p)$ is homogeneous of degree 1.
\item Therefore $t(p) p -F^2(t(p))$ is homogeneous of degree 2.
\item Therefore $G(p)$ is homogeneous of degree 1.
\end{itemize}
Thus from a Finsler \index{Finsler!function} function $F(x,t)$ we can
always construct a co--Finsler \index{co--Finsler~geometry} function
$G(x,p)$ which is homogeneous of degree 1 on the cotangent bundle.

{From} the way the proof is set up it is clearly reversible --- if you
are given a co--Finsler \index{co--Finsler~geometry} function $G(x,p)$
on the cotangent bundle this provides a \emph{natural} way of
extracting the corresponding Finsler function:
\begin{equation}
F^2(x,t) = t \; p(t) - G^2(x,p(t)).
\end{equation}

%+++++++++++++++++++++++++++++++++++++++++++++++++++++++++++++++++
\subsection{Connection with the quasi-particle PDE analysis}
%+++++++++++++++++++++++++++++++++++++++++++++++++++++++++++++++++

{From} the PDE-based analysis we obtain the second-order system of PDEs
\begin{equation}
\partial_a \left( f^{ab}{}_{\A\B}\; \partial_b \theta^B \right)
+ \hbox{lower order terms} = 0.
\label{E:system}
\end{equation}
We are now generalizing in the obvious manner to any arbitrary number
$n$ of interacting BECs, but the analysis is even more general than
that --- it applies to any field-theory normal-mode analysis that
arises from a wide class of Lagrangian based systems~\cite{normal,normal2}.

Going to the eikonal\index{eikonal} approximation this becomes
\begin{equation}
f^{ab}_{\A\B} \; p_a p_b \; \epsilon^B + \hbox{lower-order terms} = 0,
\end{equation}
which leads (neglecting lower order terms for now) to the
Fresnel-like\index{Fresnel~equation} equation
\begin{equation}
\det[ f^{ab}_{\A\B} \; p_a p_b] = 0.
\end{equation}
But by expanding the $n\times n$ determinant ($n$ is the number of
fields, not the dimension of spacetime) we have
\begin{equation}
\det[ f^{ab}_{\A\B} \; p_a p_b]  = Q^{abcd\dots} \; p_a p_b p_c p_d \dots
\end{equation}
where if there are $n$ fields there will be $2n$ factors of $p$. 

Now define
\begin{equation}
Q(x,p) =  Q^{abcd\dots} \; p_a p_b p_c p_d \dots,
\end{equation}
and
\begin{equation}
G(x,p) = \sqrt[2n]{Q(x,p)} = [Q(z,p)]^{1/(2n)},
\end{equation}
then
\begin{itemize}
\item $Q(x,p)$ is homogeneous of degree $2n$.
\item $G(x,p)$ is homogeneous of degree 1, and hence is a co--Finsler
  \index{co--Finsler~geometry} function.
\item We can now Legendre transform $G\to F$, providing a chain
\begin{equation}
Q(x,p) \to  G(x,p) \to F(x,t).
\end{equation}
Can this route be reversed? 
\end{itemize}
\emph{Step 1:} We can always reverse $F(x,t)\to G(x,p)$ by Legendre transformation.

\noindent
\emph{Step 2:} We can always define
\begin{equation}
g^{ab} (x,p) = {1\over2} {\partial\over\partial p_a} {\partial\over\partial p_b} [G(x,p)^2],
\end{equation}
this is homogeneous of degree 0, but is generically not smooth at
$p=0$.

In fact, \emph{if} $g^{ab}(x,p)$ is smooth at $p=0$ then there exits a limit
\begin{equation}
g^{ab}(x,p\to 0)= \bar g^{ab}(x),
\end{equation}
but since $g^{ab}(x,p)$ is homogeneous of degree 0 this implies
\begin{equation}
g^{ab}(x,p) = \bar g^{ab}(x) \qquad [\forall p],
\end{equation}
and so the geometry simplifies Finsler $\to$ Riemann.

This observation suggests the following definition.

\noindent\emph{Definition:} A co--Finsler \index{co--Finsler~geometry}
function $G(x,p)$ is $2n$-smooth iff the limit
\begin{equation}
{1\over(2n)!} \lim_{p\to0} \left({\partial\over\partial p}\right)^{2n} G(x,p)^{2n} =
\bar Q^{abcd\dots}
\end{equation}
exists independent of the direction $p$ in which you approach zero.

\noindent\emph{Lemma:} If $G(x,p)$ is $2n$-smooth then
\begin{equation}
G(x,p)^{2n} = \bar Q^{abcd\dots} \; p_a p_b p_c p_d \dots,
\end{equation}
and indeed
\begin{equation}
G(x,p) = \sqrt[2n]{\bar Q^{abcd\dots} \; p_a p_b p_c p_d \dots}.
\end{equation}

\noindent
\emph{Proof:} $G^{2n}$ is homogeneous of degree $2n$, so
$(\partial/\partial p)^{2n} G^{2n}$ is homogeneous of degree 0.
Therefore if the limit
\begin{equation}
{1\over(2n)!} \lim_{p\to0} \left({\partial\over\partial p}\right)^{2n} G(x,p)^{2n} =
\bar Q^{abcd\dots}
\end{equation}
exists, it follows that
\begin{equation}
{1\over(2n)!}  \left({\partial\over\partial p}\right)^{2n} G(x,p)^{2n} =
\bar Q^{abcd\dots} \qquad [\forall p],
\end{equation}
and so the result follows.

\noindent
\emph{Special case $n=1$:} If $G(x,p)$ is 2-smooth then
\begin{equation}
{1\over2}  {\partial^2\over\partial p^a \;\partial p_b } G(x,p)^{2} =
\bar Q^{ab} = g^{ab}(x,p),
\end{equation}
and co--Finsler \index{co--Finsler~geometry} $\to$ Riemann.

\noindent 
These observations have a number of implications:
\begin{itemize}
\item For all those co--Finsler \index{co--Finsler~geometry} functions
  that are $2n$ smooth we \emph{can} recover the tensor
  $Q^{abcd\dots}$.
\item Not all co--Finsler \index{co--Finsler~geometry} functions are
  $2n$ smooth, and for those functions we \emph{cannot} extract
  $Q^{abcd\dots}$ in any meaningful way.
\item But those specific co--Finsler \index{co--Finsler~geometry}
  functions that arise from the leading symbol of a 2nd-order system
  of PDEs are naturally $2n$-smooth, and so for the specific
  co--Finsler \index{co--Finsler~geometry} structures we are physically
  interested in
\begin{equation}
Q(x,p) \leftrightarrow G(x,p) \leftrightarrow F(x,t).
\end{equation}
\item Therefore, in the physically interesting case the Finsler
  function $F(x,t)$ encodes all the information present in
  $Q^{abcd\dots}$.
\end{itemize}

\noindent
\emph{Special case $n=2$:} For two fields (appropriate for our 2-BEC
\index{2-BEC~system} system), we can follow the chain
\begin{equation}
\mathbf{f}^{ab} \to Q(x,p) \leftrightarrow G(x,p) \leftrightarrow F(x,t)
\end{equation}
to formally write
\begin{equation}
ds^4 = g_{abcd} \; dx^q dx^b dx^c dx^d,
\end{equation}
or
\begin{equation}
ds = \sqrt[4]{g_{abcd} \; dx^q dx^b dx^c dx^d}.
\end{equation}
This is one of the ``more general'' cases Riemann alludes to in his
inaugural lecture of 1854~\cite{Riemann}.

This discussion makes it clear that the general geometry in our 2-BEC
\index{2-BEC~system} system is a 4-smooth Finsler geometry. It is only
for certain special cases that the Finsler geometry specializes first
to ``multi-metric'' and then to ``mono-metric'' Riemannian geometries.

%+++++++++++++++++++++++++++++++++++++++++++++++++++++++++++++++++
\subsection{Lorentzian signature Finsler geometries}
%+++++++++++++++++++++++++++++++++++++++++++++++++++++++++++++++++

The distinction between Finsler and pseudo--Finsler
\index{pseudo--Finsler~spacetime} geometries has to do with the
distinction between elliptic \index{elliptic~system} and hyperbolic
\index{hyperbolic~system} PDEs. Elliptic \index{elliptic~system} PDEs
lead to ordinary Finsler geometries, hyperbolic PDEs lead to
\emph{pseudo}--Finsler geometries.

Remember that in special relativity we typically define
\begin{equation}
d_\gamma(x,y) = \int_x^y \sqrt{ g_{ab} (dx^a/d\tau) (dx^b/d\tau)} d\tau,
\end{equation}
then
\begin{itemize}
\item $d_\gamma(x,y) \in I\!\!R^+$ for spacelike paths;
\item $d_\gamma(x,y) =0$ for null paths;
\item $d_\gamma(x,y) \in I\!\!I^+$ for timelike paths;
\end{itemize}
The point is that even in special relativity\index{special~relativity}
(and by implication in general relativity)\index{general~relativity}
``distances'' do not have to be real numbers. This is why physicists
deal with \emph{pseudo}--Riemannian [Lorentzian] geometries, not
(strictly speaking) Riemannian geometries.

To see how this generalizes in a Finsler situation let us first
consider a co--Finsler \index{co--Finsler~geometry} structure that is
multi-metric, that is:
\begin{equation}
Q(x,p) = \Pi_{i=1}^n (g^{ab}_i p_a p_b),
\end{equation}
where each one of these $n$ factors contains a Lorentzian signature
matrix and so can pass through zero.  Then
\begin{equation}
G(x,p) = \sqrt[2n]{ \Pi_{i=1}^n (g^{ab}_i p_a p_b) },
\end{equation}
and 
\begin{equation}
G(x,p) \in \exp\left({i\pi \ell\over2n}\right) \; I\!\!R^+,
\end{equation}
where
\begin{itemize}
\item $\ell=0 \to G(x,p)\in  I\!\!R^+ \to$ outside all $n$ signal cones;
\item $\ell=n \to G(x,p)\in  I\!\!I^+ \to$ inside all $n$ signal cones.
\end{itemize}
So we can now define
\begin{itemize}
\item Spacelike $\leftrightarrow$ outside all $n$ signal cones $\leftrightarrow$ $G$ real;
\item Null $\leftrightarrow$ on any one of the $n$ signal cones $\leftrightarrow$ $G$ zero;
\item Timelike $\leftrightarrow$ inside all $n$ signal cones $\leftrightarrow$ $G$ imaginary;
\item plus the various ``intermediate'' cases:
\begin{equation}
\hbox{``intermediate''} \leftrightarrow  \hbox{inside $\ell$ of $n$ signal cones} 
\leftrightarrow G\in i^{\ell/n} \times I\!\!R^+.
\end{equation}
\end{itemize}
Now this basic idea survives even if we do not have a multi-metric
theory. The condition $Q(x,p)=0$ defines a polynomial of degree $2n$,
and so defines $n$ nested sheets (possibly crossing in places).
Compare with Courant and Hilbert's discussion of the Monge
cone~\cite{Courant}.

That is:
\begin{eqnarray*}
Q(x,p)=0
&\Leftrightarrow&
Q(x,(E,\vec p)) =0;
\\
&\Leftrightarrow& 
\hbox{polynomial of degree $2n$ in $E$ for any fixed $\vec p$;}
\\
&\Leftrightarrow& 
\hbox{in each direction $\exists$ $2n$ roots in $E$;}
\\
&\Leftrightarrow& 
\hbox{corresponds to $n$ [topological] cones.}
\end{eqnarray*}
(These are topological cones, not geometrical cones, and the roots
might happen to be degenerate.)

\noindent\emph{Question:} Should we be worried by the fact that the
co-metric $g^{ab}$ is singular on the signal cone? (In fact on all $n$
of the signal cones.) Not really.  We have
\begin{equation}
G(x,p) = \sqrt[2n]{\bar Q^{abcd\dots} \; p_a p_b p_c p_d \dots},
\end{equation}
so
\begin{equation}
g^{ab}(x,p) = {1\over2}  
{\partial^2\over\partial p^a \;\partial p_b } \left( \sqrt[n]{Q(x,p)} \right)
=
{1\over2n} {\partial\over\partial p_b } 
\left\{ Q^{{1\over n}-1} \; Q^{abcd\dots} \; p_b p_c p_d \dots \right\},
\end{equation}
whence
\begin{eqnarray}
g^{ab}(x,p) &=& {1\over2n} 
Q^{{1\over n}-1} \; Q^{abcd\dots} \; p_c p_d \dots 
\\
&&
+ {1\over2n} \left({1\over n}-1\right) Q^{{1\over n}-2} \left[ Q^{acde\dots} \; p_c p_d p_e \dots \right]\; \left[  Q^{bfgh\dots} \; p_f p_g p_h \dots \right],
\nonumber
\end{eqnarray}
which we can write as
\begin{eqnarray}
g^{ab}(x,p) &=& {1\over2n} 
Q^{-(n-1)/n} \; Q^{abcd\dots} \; p_c p_d \dots 
\\
&&
- {1\over2n} {n-1\over n} Q^{-(2n-1)/n} 
\left[ Q^{acde\dots} \; p_c p_d p_e \dots \right]\; \left[  Q^{bfgh\dots} \; p_f p_g p_h \dots \right].
\nonumber
\end{eqnarray}
Yes, this naively looks like it's singular on the signal cone
where $Q(x,p)=0$.  But no, this is not a problem: Consider
\begin{equation}
g^{ab} p_a p_b = {1\over2n}  Q^{-(2n-1)/n} Q - {1\over2n} {n-1\over n} Q^{-(2n-1)/n} Q^2,
\end{equation}
then 
\begin{equation}
  g^{ab} p_a p_b = {1\over2n}\left(1-{n-1\over n}\right) Q^{1/n} = {1\over2n^2} \; Q^{1/n} = 0,
\end{equation}
and this quantity is definitely non-singular.

%+++++++++++++++++++++++++++++++++++++++++++++++++++++++++++++++++
\subsection{Summary} 
%+++++++++++++++++++++++++++++++++++++++++++++++++++++++++++++++++

In short: 
\begin{itemize}
\item pseudo--Finsler \index{pseudo--Finsler~spacetime} functions
  arise naturally from the leading symbol of hyperbolic
  \index{hyperbolic~system} systems of PDEs;
\item pseudo--Finsler \index{pseudo--Finsler~spacetime} geometries
  provide the natural ``geometric'' interpretation of a
  multi-component PDE before fine tuning;
\item In particular the natural geometric interpretation of our 2-BEC
  \index{2-BEC~system} model (before fine tuning) is as a 4-smooth
  pseudo--Finsler \index{pseudo--Finsler~spacetime} geometry.
\end{itemize}

%%%%%%%%%%%%%%%%%%%%%%%%%%%%%%%%%%%%%%%%%%%%%%%%%%%

\section{Some matrix identities}     

%%%%%%%%%%%%%%%%%%%%%%%%%%%%%%%%%%%%%%%%%%%%%%%%%%%

To simplify the flow of argument in the body of the paper, here we
collect a few basic results on $2\times 2$ matrices that are used in
our analysis.

%+++++++++++++++++++++++++++++++++++++++++++++++++++++++++++++++++
\subsection{Determinants}
%+++++++++++++++++++++++++++++++++++++++++++++++++++++++++++++++++

\noindent{\bf Theorem:} For any two $2\times 2$ matrix
$A$:
\begin{equation}
\label{E:A-det2}
\det(A)=  {1\over2} \left\{ \tr[A]^2-\tr[A^2]\right\} . 
\end{equation}
This is best proved by simply noting
\begin{equation}
\det (A) = \lambda_1 \lambda_2 = 
{1\over2}\left[ (\lambda_1 +\lambda_2)^2 - (\lambda_1^2 +\lambda_2^2) \right]
= {1\over2} \left\{ \tr[A]^2-\tr[A^2]\right\} . 
\end{equation}
If we now define $2\times2$ ``trace reversal'' (in a manner
reminiscent of standard GR) by
\begin{equation}
\bar A = A - \tr[A]\;\mathbf{I};  \qquad   \bar{\!\!\bar A} = A;
\end{equation}
then this looks even simpler
\begin{equation}
\label{E:A-det23}
\det(A)=  -{1\over2} \tr[A\;\bar A] = \det(\bar A). 
\end{equation}

\noindent
A simple implication is now:\\
\noindent{\bf Theorem:} For any two $2\times 2$ matrices
$A$ and $B$:
\begin{equation}
\label{E:A-2-matrices2}
\det(A+\lambda \; B) = \det(A) + \lambda\;\left\{ \tr[A] \tr[B] - \tr[  A\;B]
\right\}  + \lambda^2\;\det(B).
\end{equation}
which we can also write as
\begin{equation}
\label{E:A-2-matrices}
\det(A+\lambda \; B) = \det(A) - \lambda\; \tr[  A\;\bar B]
+ \lambda^2\;\det(B).
\end{equation}
Note that $\tr[A\;\bar B] = \tr[\bar A\; B]$.

%+++++++++++++++++++++++++++++++++++++++++++++++++++++++++++++++++
\subsection{Hamilton--Cayley theorems}
%+++++++++++++++++++++++++++++++++++++++++++++++++++++++++++++++++

\noindent{\bf Theorem:} For any two $2\times 2$ matrix
$A$:
\begin{equation}
A^{-1} = \frac{\tr[A] \;\; \mathbf{I} - A }{\det[A]} = -{\bar A\over\det{[\bar A]}} .
\end{equation}

\noindent{\bf Theorem:} For any two $2\times 2$ matrix
$A$:
\begin{equation}
A^{1/2} = \;\pm \left\{
\frac{A\pm\sqrt{\det A} \;\; \mathbf{I} }{\sqrt{\tr[A]\pm 2  \sqrt{\det A}}}
\right\} .
\end{equation}
%

%%%%%%%%%%%%%%%%%%%%%%%%%%%%%%%%%%%%%%%%%%%%%%%%%%%

%%%%%%%%%%%%%%%%%%%%%%%%%%%%%%%%%%%%%%%%%%%%%%%%%%%
																	     
%%%%%%%%%%%%%%%%%%%%%%%%%%%%%%%%%%%%%%%%%%%%%%%%%%%
\input{reference}

%%%%%%%%%%%%%%%%%%%%%%%%%%%%%%%%%%%%%%%%%%%%%%%%%%%

%%%%%%%%%%%%%%%%%%%%%%%%%%%%%%%%%%%%%%%%%%%%%%%%%%%
\printindex
%%%%%%%%%%%%%%%%%%%%%%%%%%%%%%%%%%%%%%%%%%%%%%%%%%%

%%%%%%%%%%%%%%%%%%%%%%%%%%%%%%%%%%%%%%%%%%%%%%%%%%%
\setcounter{tocdepth}{2}
\tableofcontents %%% for debugging purposes only
%%%%%%%%%%%%%%%%%%%%%%%%%%%%%%%%%%%%%%%%%%%%%%%%%%%

\end{document}

%% file: energylevel.pstex_t
\begin{picture}(0,0)%
\includegraphics{energylevel.pstex}%
\end{picture}%
\setlength{\unitlength}{2368sp}%
\begingroup\makeatletter\ifx\SetFigFont\undefined%
\gdef\SetFigFont#1#2#3#4#5{%
  \reset@font\fontsize{#1}{#2pt}%
  \fontfamily{#3}\fontseries{#4}\fontshape{#5}%
  \selectfont}%
\fi\endgroup%
\begin{picture}(7807,6080)(-599,-7319)
\put(-599,-1636){\makebox(0,0)[lb]{\smash{{\SetFigFont{7}{8.4}{\rmdefault}{\mddefault}{\updefault}{\color[rgb]{0,0,0}$\vert e \rangle$}%
}}}}
\put(4951,-5761){\makebox(0,0)[lb]{\smash{{\SetFigFont{7}{8.4}{\familydefault}{\mddefault}{\updefault}{\color[rgb]{0,0,0}$F=1$}%
}}}}
\put(4951,-4711){\makebox(0,0)[lb]{\smash{{\SetFigFont{7}{8.4}{\familydefault}{\mddefault}{\updefault}{\color[rgb]{0,0,0}$F=2$}%
}}}}
\put(6301,-5236){\makebox(0,0)[lb]{\smash{{\SetFigFont{7}{8.4}{\familydefault}{\mddefault}{\updefault}{\color[rgb]{0,0,0}$6.8 GHz$}%
}}}}
\put(6301,-1636){\makebox(0,0)[lb]{\smash{{\SetFigFont{7}{8.4}{\familydefault}{\mddefault}{\updefault}{\color[rgb]{0,0,0}$800 MHz$}%
}}}}
\put(4951,-1411){\makebox(0,0)[lb]{\smash{{\SetFigFont{7}{8.4}{\familydefault}{\mddefault}{\updefault}{\color[rgb]{0,0,0}$F'=2$}%
}}}}
\put(  1,-2086){\makebox(0,0)[lb]{\smash{{\SetFigFont{7}{8.4}{\familydefault}{\mddefault}{\updefault}{\color[rgb]{0,0,0}$\Delta$}%
}}}}
\put(4951,-1861){\makebox(0,0)[lb]{\smash{{\SetFigFont{7}{8.4}{\familydefault}{\mddefault}{\updefault}{\color[rgb]{0,0,0}$F'=1$}%
}}}}
\put(451,-3661){\makebox(0,0)[lb]{\smash{{\SetFigFont{7}{8.4}{\familydefault}{\mddefault}{\updefault}{\color[rgb]{0,0,0}$\Omega_1$}%
}}}}
\put(2101,-3361){\makebox(0,0)[lb]{\smash{{\SetFigFont{7}{8.4}{\familydefault}{\mddefault}{\updefault}{\color[rgb]{0,0,0}$\Omega_2$}%
}}}}
\put(-599,-5536){\makebox(0,0)[lb]{\smash{{\SetFigFont{7}{8.4}{\familydefault}{\mddefault}{\updefault}{\color[rgb]{0,0,0}$\vert g \rangle$}%
}}}}
\put(151,-6661){\makebox(0,0)[lb]{\smash{{\SetFigFont{7}{8.4}{\familydefault}{\mddefault}{\updefault}{\color[rgb]{0,0,0}$-2$}%
}}}}
\put(1051,-6661){\makebox(0,0)[lb]{\smash{{\SetFigFont{7}{8.4}{\familydefault}{\mddefault}{\updefault}{\color[rgb]{0,0,0}$-1$}%
}}}}
\put(1951,-6661){\makebox(0,0)[lb]{\smash{{\SetFigFont{7}{8.4}{\familydefault}{\mddefault}{\updefault}{\color[rgb]{0,0,0}$0$}%
}}}}
\put(2851,-6661){\makebox(0,0)[lb]{\smash{{\SetFigFont{7}{8.4}{\familydefault}{\mddefault}{\updefault}{\color[rgb]{0,0,0}$+1$}%
}}}}
\put(3751,-6661){\makebox(0,0)[lb]{\smash{{\SetFigFont{7}{8.4}{\familydefault}{\mddefault}{\updefault}{\color[rgb]{0,0,0}$+2$}%
}}}}
\put(1951,-7261){\makebox(0,0)[lb]{\smash{{\SetFigFont{7}{8.4}{\familydefault}{\mddefault}{\updefault}{\color[rgb]{0,0,0}$m_F$}%
}}}}
\end{picture}%

%% file: geometry.pstex_t
\begin{picture}(0,0)%
\includegraphics{geometry.pstex}%
\end{picture}%
\setlength{\unitlength}{3394sp}%
\begingroup\makeatletter\ifx\SetFigFont\undefined%
\gdef\SetFigFont#1#2#3#4#5{%
  \reset@font\fontsize{#1}{#2pt}%
  \fontfamily{#3}\fontseries{#4}\fontshape{#5}%
  \selectfont}%
\fi\endgroup%
\begin{picture}(6399,3999)(-7886,-3523)
\put(-4799,-2911){\makebox(0,0)[rb]{\smash{{\SetFigFont{7}{8.4}{\sfdefault}{\mddefault}{\updefault}{\color[rgb]{0.258,0.258,0.258}$d_{\A}\tilde{U}_{\A\A}=d_{\B}\tilde{U}_{\B\B}$}%
}}}}
\put(-1874,-2311){\makebox(0,0)[rb]{\smash{{\SetFigFont{7}{8.4}{\sfdefault}{\mddefault}{\updefault}{\color[rgb]{0,0,0}$\vec{v}_{\A 0}=\vec{v}_{\B 0}$}%
}}}}
\put(-4649,-1616){\rotatebox{270.0}{\makebox(0,0)[lb]{\smash{{\SetFigFont{7}{8.4}{\sfdefault}{\mddefault}{\updefault}{\color[rgb]{0.258,0.258,0.258}special case, $\Xi$= const.}%
}}}}}
\put(-7799,-3286){\makebox(0,0)[lb]{\smash{{\SetFigFont{9}{10.8}{\sfdefault}{\mddefault}{\updefault}{\color[rgb]{0,0,0}mono-metric}%
}}}}
\put(-7799,-1936){\makebox(0,0)[lb]{\smash{{\SetFigFont{9}{10.8}{\sfdefault}{\mddefault}{\updefault}{\color[rgb]{0,0,0}bi-metric}%
}}}}
\put(-7799,-1036){\makebox(0,0)[lb]{\smash{{\SetFigFont{9}{10.8}{\sfdefault}{\mddefault}{\updefault}{\color[rgb]{0,0,0}pseudo-Finsler}%
}}}}
\put(-7799,-661){\makebox(0,0)[lb]{\smash{{\SetFigFont{9}{10.8}{\sfdefault}{\mddefault}{\updefault}{\color[rgb]{0,0,0}$\mathbf{f}^{ab}=\left[ \mathbf{f}^{\T} \right]^{ba} $}%
}}}}
\put(-5324,-811){\makebox(0,0)[lb]{\smash{{\SetFigFont{9}{10.8}{\sfdefault}{\mddefault}{\updefault}{\color[rgb]{0,0,0}$\vec{v}_{\A 0}=\vec{v}_{\B 0}$}%
}}}}
\put(-2774,-1036){\makebox(0,0)[lb]{\smash{{\SetFigFont{9}{10.8}{\sfdefault}{\mddefault}{\updefault}{\color[rgb]{0,0,0}$\mathbf{f}^{ab}=\mathbf{f}^{ba} $}%
}}}}
\put(-4274,-1561){\makebox(0,0)[rb]{\smash{{\SetFigFont{7}{8.4}{\sfdefault}{\mddefault}{\updefault}{\color[rgb]{0,0,0}$\tilde{U}_{\A\B}=0$}%
}}}}
\put(-4124,-1711){\makebox(0,0)[lb]{\smash{{\SetFigFont{7}{8.4}{\sfdefault}{\mddefault}{\updefault}{\color[rgb]{0,0,0}$d_{\A}=d_{\B}$}%
}}}}
\put(-6524,-1261){\makebox(0,0)[lb]{\smash{{\SetFigFont{7}{8.4}{\sfdefault}{\mddefault}{\updefault}{\color[rgb]{0,0,0}$\tilde{U}_{\A\B}=0$}%
}}}}
\put(-3299,-61){\makebox(0,0)[lb]{\smash{{\SetFigFont{10}{12.0}{\sfdefault}{\mddefault}{\updefault}{\color[rgb]{0,0,0}Emergent Geometry}%
}}}}
\put(-2774,-736){\makebox(0,0)[lb]{\smash{{\SetFigFont{9}{10.8}{\sfdefault}{\mddefault}{\updefault}{\color[rgb]{0,0,0}$\mathbf{f}^{ab}=\left[ \mathbf{f}^{\T} \right]^{ab} $}%
}}}}
\put(-1874,-2761){\makebox(0,0)[rb]{\smash{{\SetFigFont{7}{8.4}{\sfdefault}{\mddefault}{\updefault}{\color[rgb]{0,0,0}$\tilde{U}_{\A\B}=0$}%
}}}}
\put(-1874,-2986){\makebox(0,0)[rb]{\smash{{\SetFigFont{7}{8.4}{\sfdefault}{\mddefault}{\updefault}{\color[rgb]{0,0,0}$ \tilde{U}_{\A\A}= \tilde{U}_{\B\B}$}%
}}}}
\put(-1874,-2536){\makebox(0,0)[rb]{\smash{{\SetFigFont{7}{8.4}{\sfdefault}{\mddefault}{\updefault}{\color[rgb]{0,0,0}$d_{\A}=d_{\B}$}%
}}}}
\put(-7781, 32){\rotatebox{270.0}{\makebox(0,0)[lb]{\smash{{\SetFigFont{6}{7.2}{\sfdefault}{\mddefault}{\updefault}{\color[rgb]{0,0,0}tuning}%
}}}}}
\put(-7583,296){\makebox(0,0)[lb]{\smash{{\SetFigFont{6}{7.2}{\sfdefault}{\mddefault}{\updefault}{\color[rgb]{0,0,0}geometry}%
}}}}
\end{picture}%

%% file: UV_Physics.pstex_t
\begin{picture}(0,0)%
\includegraphics{UV_Physics.pstex}%
\end{picture}%
\setlength{\unitlength}{3394sp}%
\begingroup\makeatletter\ifx\SetFigFont\undefined%
\gdef\SetFigFont#1#2#3#4#5{%
  \reset@font\fontsize{#1}{#2pt}%
  \fontfamily{#3}\fontseries{#4}\fontshape{#5}%
  \selectfont}%
\fi\endgroup%
\begin{picture}(6399,3999)(2914,-2695)
\put(3526, 14){\rotatebox{90.0}{\makebox(0,0)[lb]{\smash{{\SetFigFont{7}{8.4}{\sfdefault}{\mddefault}{\updefault}{\color[rgb]{0.258,0.258,0.258}energy}%
}}}}}
\put(6526,-961){\makebox(0,0)[lb]{\smash{{\SetFigFont{9}{10.8}{\sfdefault}{\mddefault}{\updefault}{\color[rgb]{0,0,0}Lorentz Violation}%
}}}}
\put(3076,-1561){\rotatebox{90.0}{\makebox(0,0)[lb]{\smash{{\SetFigFont{7}{8.4}{\rmdefault}{\bfdefault}{\updefault}{\color[rgb]{0,0,0}quantum-pressure}%
}}}}}
\put(3376,-436){\rotatebox{90.0}{\makebox(0,0)[lb]{\smash{{\SetFigFont{7}{8.4}{\rmdefault}{\bfdefault}{\updefault}{\color[rgb]{0,0,0}included}%
}}}}}
\put(3376,-2011){\rotatebox{90.0}{\makebox(0,0)[lb]{\smash{{\SetFigFont{7}{8.4}{\rmdefault}{\bfdefault}{\updefault}{\color[rgb]{0,0,0} neglected}%
}}}}}
\put(4051,-1936){\makebox(0,0)[lb]{\smash{{\SetFigFont{9}{10.8}{\sfdefault}{\mddefault}{\updefault}{\color[rgb]{0,0,0}disp. rel.}%
}}}}
\put(7276,-1786){\makebox(0,0)[rb]{\smash{{\SetFigFont{8}{9.6}{\sfdefault}{\mddefault}{\updefault}{\color[rgb]{0,0,0}1 disp. rel.}%
}}}}
\put(6526,-1336){\makebox(0,0)[lb]{\smash{{\SetFigFont{9}{10.8}{\sfdefault}{\mddefault}{\updefault}{\color[rgb]{0,0,0}Lorentz Invariance}%
}}}}
\put(8851,-661){\makebox(0,0)[lb]{\smash{{\SetFigFont{8}{9.6}{\sfdefault}{\mddefault}{\updefault}{\color[rgb]{0,0,0}$\hat{\Xi}$}%
}}}}
\put(8851,-1711){\makebox(0,0)[lb]{\smash{{\SetFigFont{8}{9.6}{\sfdefault}{\mddefault}{\updefault}{\color[rgb]{0,0,0}$\Xi$}%
}}}}
\put(3826,-2086){\rotatebox{90.0}{\makebox(0,0)[lb]{\smash{{\SetFigFont{9}{10.8}{\sfdefault}{\mddefault}{\updefault}{\color[rgb]{0,0,0}phononic}%
}}}}}
\put(3826,-1036){\rotatebox{90.0}{\makebox(0,0)[lb]{\smash{{\SetFigFont{9}{10.8}{\sfdefault}{\mddefault}{\updefault}{\color[rgb]{0,0,0}trans-phonic}%
}}}}}
\put(3751,-2536){\makebox(0,0)[lb]{\smash{{\SetFigFont{9}{10.8}{\sfdefault}{\mddefault}{\updefault}{\color[rgb]{0,0,0}pseudo-Finsler}%
}}}}
\put(5251,-2536){\makebox(0,0)[lb]{\smash{{\SetFigFont{9}{10.8}{\sfdefault}{\mddefault}{\updefault}{\color[rgb]{0,0,0}bi-metric}%
}}}}
\put(6601,-2536){\makebox(0,0)[lb]{\smash{{\SetFigFont{9}{10.8}{\sfdefault}{\mddefault}{\updefault}{\color[rgb]{0,0,0}mono-metric}%
}}}}
\put(4051,-1711){\makebox(0,0)[lb]{\smash{{\SetFigFont{8}{9.6}{\sfdefault}{\mddefault}{\updefault}{\color[rgb]{0,0,0}2 coupled}%
}}}}
\put(7051,764){\makebox(0,0)[lb]{\smash{{\SetFigFont{10}{12.0}{\sfdefault}{\mddefault}{\updefault}{\color[rgb]{0,0,0}High-energy excitations}%
}}}}
\put(6151,-1711){\makebox(0,0)[rb]{\smash{{\SetFigFont{8}{9.6}{\sfdefault}{\mddefault}{\updefault}{\color[rgb]{0,0,0}2 independent}%
}}}}
\put(5251,-1936){\makebox(0,0)[lb]{\smash{{\SetFigFont{9}{10.8}{\sfdefault}{\mddefault}{\updefault}{\color[rgb]{0,0,0}disp. rel.}%
}}}}
\put(3070,854){\rotatebox{90.0}{\makebox(0,0)[lb]{\smash{{\SetFigFont{6}{7.2}{\sfdefault}{\mddefault}{\updefault}{\color[rgb]{0,0,0}energy}%
}}}}}
\put(3220,479){\makebox(0,0)[lb]{\smash{{\SetFigFont{6}{7.2}{\sfdefault}{\mddefault}{\updefault}{\color[rgb]{0,0,0}geometry}%
}}}}
\put(8701,-2461){\makebox(0,0)[lb]{\smash{{\SetFigFont{7}{8.4}{\sfdefault}{\mddefault}{\updefault}{\color[rgb]{0,0,0}tuning}%
}}}}
\end{picture}%

%% file: reference.tex
%%%%%%%%%%%%%%%%%%%%%%%% referenc.tex %%%%%%%%%%%%%%%%%%%%%%%%%%%%%%
% sample references
% "physics"
%
% Use this file as a template for your own input.
%
%%%%%%%%%%%%%%%%%%%%%%%% Springer-Verlag %%%%%%%%%%%%%%%%%%%%%%%%%%
%
% BibTeX users please use
% \bibliographystyle{}
% \bibliography{}
%
% Non-BibTeX users please use